\documentclass{aa}
\usepackage{graphicx}
\usepackage{txfonts}

\begin{document}

\def\ltsima{$\; \buildrel < \over \sim \;$}
\def\simlt{\lower.5ex\hbox{\ltsima}}
\def\gtsima{$\; \buildrel > \over \sim \;$}
\def\simgt{\lower.5ex\hbox{\gtsima}}

\title{Efficient differential Fourier-transform spectrometer \\ for precision Sunyaev-Zel'dovich effect measurements}

\author{Alessandro Schillaci\inst{1}, Giuseppe D' Alessandro\inst{1}, Paolo de Bernardis\inst{1}, \\
Silvia Masi\inst{1}, Camila Paiva Novaes \inst{2}, Massimo Gervasi \inst{3}, Mario Zannoni\inst{3}}

\institute{ Dipartimento di Fisica, Universit\`{a} di Roma ``La Sapienza", Roma, Italy
    \and Divis\~ao de Astrof\'isica, Instituto Nacional de Pesquisas Espaciais, S\~ao Jos\'e dos Campos, SP, Brazil
    \and Dipartimento di Fisica G. Occhialini, Universit\'a Milano Bicocca, Milano, Italy}

\offprints{alessandro.schillaci@roma1.infn.it}

\date{Submitted: Jan. 12$^{th}$, 2014 ; revised: Apr. 4$^{th}$, 2014; Accepted: Apr. 11$^{th}$, 2014}

{ \abstract {Precision measurements of the Sunyaev-Zel'dovich effect in clusters of galaxies require excellent rejection of common-mode signals and wide frequency coverage.}{We describe an imaging, efficient, differential Fourier transform spectrometer (FTS), optimized for measurements of faint brightness gradients at millimeter wavelengths.}{Our instrument is based on a Martin-Puplett interferometer (MPI) configuration. We combined two MPIs working synchronously to use the whole input power. In our implementation the observed sky field is divided into two halves along the meridian, and each half-field corresponds to one of the two input ports of the MPI. In this way, each detector in the FTS focal planes measures the difference in brightness between two sky pixels, symmetrically located with respect to the meridian. Exploiting the high common-mode rejection of the MPI, we can measure low sky brightness gradients over a high isotropic background.} {The instrument works in the range $\sim$ 1$-$20 cm$^{-1}$ (30$-$600 GHz), has a maximum spectral resolution $1/(2 \ OPD) = 0.063 \ cm^{-1}$ (1.9 GHz), and an unvignetted throughput of 2.3 cm$^2$sr. It occupies a volume of 0.7$\times$0.7$\times$0.33 m$^3$ and has a weight of 70 kg. This design can be implemented as a cryogenic unit to be used in space, as well as a room-temperature unit working at the focus of suborbital and ground-based mm-wave telescopes. The first in-flight test of the instrument is with the OLIMPO experiment on a stratospheric balloon; a larger implementation is being prepared for the Sardinia radio telescope.}{}}


\authorrunning{A. Schillaci \emph{et al.}}
\titlerunning{Efficient DFTS for precision SZ effect measurements }
\maketitle

\section{Introduction}\label{introduction}

About 1\% of the photons of the cosmic microwave background (CMB)
that cross a rich cluster of galaxies undergo inverse Compton
scattering against the hot electrons of the intracluster plasma.
This phenomenon is called the thermal Sunyaev-Zel'dovich effect
(SZE) and is a powerful tool for astrophysical and cosmological
investigation (Sunyaev and Zeldovich \cite{SunZel70}, (Sunyaev and Zeldovich \cite{SunZel72}, Rephaeli \cite{reph95}, Birkinshaw \cite{birk99}, Carlstrom et al. \cite{Carl02}).

Most of the power of the thermal SZE is due to its very
characteristic spectrum: in the direction of a cluster of
galaxies, the SZE induces a {\sl decrease} of the brightness of
the CMB at frequencies below 217 GHz, and an {\sl increase} of the
brightness of the CMB at frequencies above 217 GHz. In the nonrelativistic limit, the spectrum of the SZE can be computed from
the Kompaneets equation (Kompaneets \cite{Komp57}) and only depends on the
Comptonization Parameter $y$,
\begin{equation}
y=\int n_{e}\sigma_{T}\frac{k_{B}T_{e}}{m_{e}c^{2}} d\ell \ ,
\end{equation}
where the integral is along the line of sight. Using the $y$
parameter, the brightness variation $\Delta I_{t}$ with respect to
the average cosmic microwave brightness $I$ can be expressed as
\begin{equation}
\frac{\Delta I_{t}}{I}= y\frac{x^{4}e^{x}}{(e^{x}-1)^{2}}\bigl[
x  coth(x/2)-4 \bigr] \ .\label{eq:tsze}
\end{equation}
Typical values of $y$ are $10^{-4} - 10^{-5}$, so the SZE
signal $\Delta I_t$ is much weaker than the average
brightness $I$ of the CMB. However, the SZE spectrum does not
depend on the distance (redshift) of the cluster of galaxies, so
clusters become important cosmological probes. Moreover, the SZE
spectrum depends linearly on both the temperature and density of
the intracluster plasma. Information from the SZE significantly complements
information obtained from X-ray brightness
(proportional to the density squared) in investigating the physics
of the clusters. The thermal SZE is only the most evident
interaction between the CMB and the plasma of the cluster. The
peculiar motion of the cluster induces the kinematic SZE, and the
nonthermal component of the plasma produces additional
distortions (Wright \cite{Wrig79}).

Our ability of measuring the SZE has recently reached maturity with the surveys
of large, dedicated ground-based telescopes (see e.g.
Reichardt et al. \cite{Reic13}, Hasselfield et al. \cite{Hass13} and references therein) at low frequencies,
and with the space-based survey of the Planck satellite (see
Planck collaboration \cite{Plan13} and references therein), which covers the whole spectral
range where the SZE is present.

All these are multiband photometric measurements. The
number of independent bands useful to this purpose is limited to a
few for ground-based observations, and was 6$-$7 in the case of
the Planck satellite. The parameters that describe the
cluster and the foregrounds can be more than seven: in principle,
more bands and/or complementary observations in other spectral
regions are needed.

Spectroscopic measurements can help very much in this situation.
They allow us to extract the weak SZE signal even in the presence
of important foreground emission, and to estimate the parameters
of the cluster by removing the degeneracy of
photometric measurements (see de Bernardis et al. \cite{debe12} and section \ref{ps} for a specific
discussion).

\begin{figure}[ht]
\centering
\includegraphics[width=8.5cm]{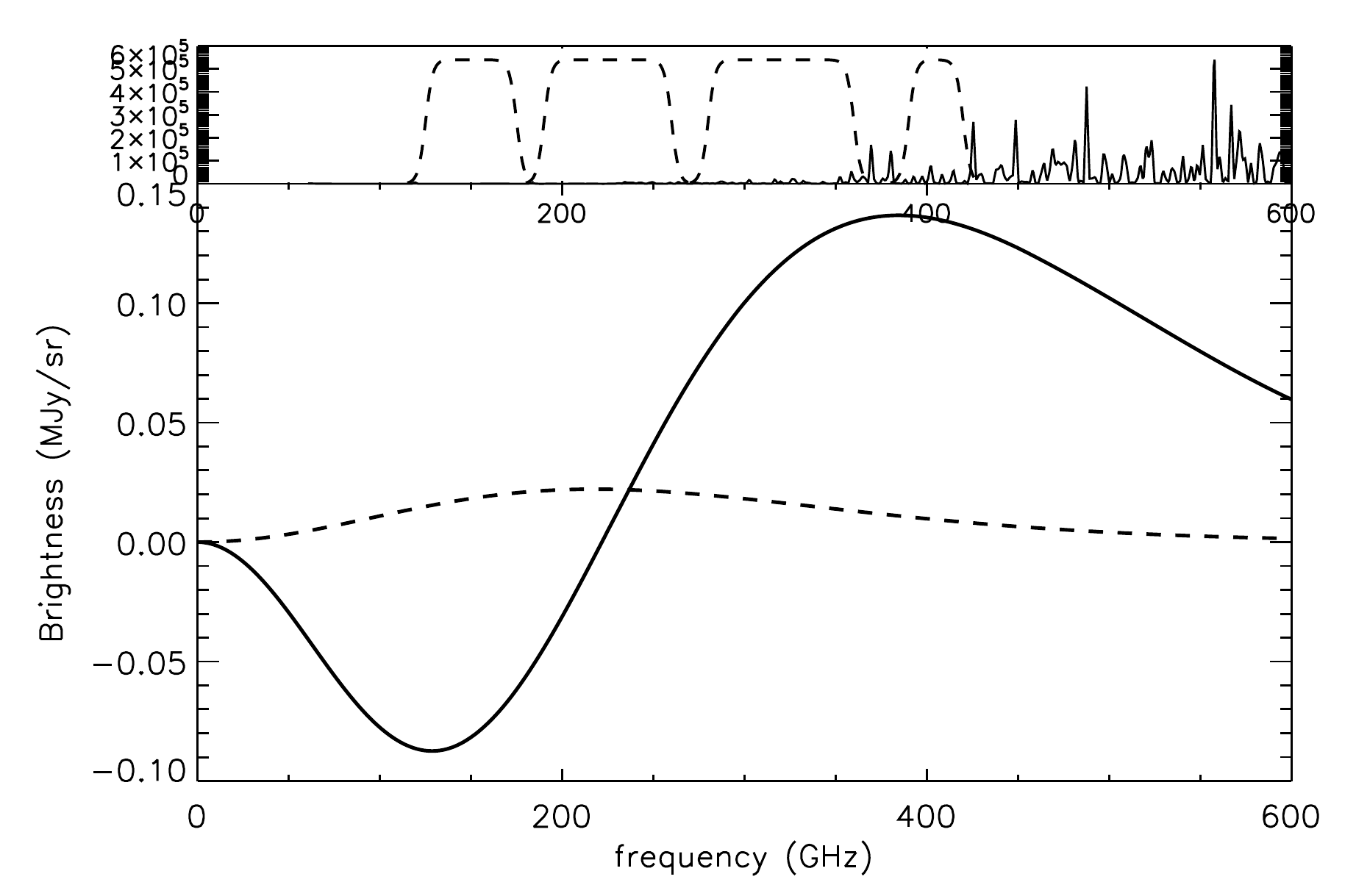}\\
\caption{{\bf Bottom:} Thermal (continuous line) and kinematic (dashed line) SZE for $y=5 \times 10^{-4}$, $v_{p}=1000 km/s$ and $T_{e}=8.2keV$, in MJy/sr.  {\bf Top:} residual atmospheric emission at stratospheric balloon altitude (continuous line, in MJy/sr as well). Four sub-bands (see text) suitable for balloon-borne obserations of the SZE are also plotted in the same panel (dashed lines).
}\label{SZE}
\end{figure}

Here we present an instrument optimized for spectroscopic
measurements of the SZE. Owing to the continuous nature of the
spectrum (see Fig. \ref{SZE}), high spectral resolution is not a
design-driver. A coarse spectral resolution ($\sim$ 0.5 cm$^{-1}$)
is more than enough to provide a sufficient number of independent
spectral data. Better resolution could be useful to assess whether there is 
contamination from molecular lines, but this is important only for low galactic latitudes.
Instead, the instrument is required to be imaging
(feeding a detector array in its focal plane), to have a wide
frequency coverage (almost two decades at mm/submm wavelengths),
and to be differential, that is, to measure the difference in
brightness between the cluster direction and the surrounding
directions, rejecting the bright isotropic background from the
CMB. The class of the Fourier transform spectrometers (FTSs), and in particular the Martin-Puplett interferometer (MPI, Martin and Puplett \cite{Mart70}), fulfills all these requirements.

Fourier transform spectrometers are imaging instruments with high spectral resolution and efficiency in the IR, with a variety of applications in chemistry, remote sensing, atmospheric studies, planetary exploration and IR astronomy. Notable space-borne implementations of differential FTSs for astrophysics and cosmology are the Far InfraRed Absolute Spectrometer (FIRAS) on-board of the COBE satellite (Mather et al. \cite{Math90}, Fixsen et al. \cite{Fixs94}) and the Spectral and Photometric Imaging Receiver (SPIRE)  FTS on-board of the Herschel satellite (Griffin et al. \cite{Grif10}). These two compare  the brightness of the sky with the brightness produced by an internal blackbody reference.

In our instrument the two input ports instead collect radiation
from two independent sky fields, so that each detector in the
focal plane of the spectrometer measures the brightness difference
between two. In this way the common-mode
signals from the instrument, the residual atmosphere and the isotropic
CMB component, are all rejected, and only brightness gradients are
measured. For this reason we use the achronym DFTS (differential Fourier transform spectrometer) for our instrument.
A previous implementation of this configuration at shorter wavelengths is the
Bear interferometer at the Canada France Hawaii Telescope (CFHT) (see e.g. Maillard \cite{Mail95}).

First light for our DFTS will be on a stratospheric balloon flight of the
OLIMPO experiment (Masi et al. \cite{Masi03}); additional configurations have
been studied for the Spectroscopic Active Galaxies And Clusters Explorer (SAGACE) satellite proposal (de Bernardis et al. \cite{debe10}), for
the millimetron satellite mission (Smirnov et al. \cite{Smir12}), and for the
Sardinia radio telescope (SRT, Grueff et al. \cite{Grue04}).

The paper is organized as follows: in section \ref{ps} we show that adding spectral measurements of the SZE to photometric measurements results in a significant enhancement of the ability to separate degenerate parameters that describe the SZE and foreground contamination.  In section  \ref{efficient} we quantitatively describe the operation of the DFTS and demostrate its efficiency. In section  \ref{design} we describe the optical design strategy for the instrument. In section \ref{mech} we describe the optical optimization  and the
opto-mechanical design. Then (section \ref{prob}) we discuss the most critical problems related to this configuration.  In section \ref{test} we describe the test and commissioning of the instrument.  We conclude in section \ref{disc}.

\section{SZE photometry and spectrometry} \label{ps}

There are many parameters that describe the SZE signal from a cluster of galaxies and the foreground emission along the same line of sight. The optical depth $\tau_T$ and the temperature $T_e$ of the plasma provide a good description of the thermal SZE. But other parameters must be considered. The cluster velocity component along the line of sight $\rm{v}_{LOS}$ produces the kinematic SZE and includes any intrinsic CMB anisotropy in the same direction (the sum of the two effects can be described by a single parameter $\Delta T_{CMB}$). At least three parameters describe the nonthermal population of electrons in the cluster plasma, which is responsible for the nonthermal SZE:  its optical depth $\tau_{nt}$,  the spectral index of the power-law spectrum of the energy of the electrons $\alpha$, and their minimum momentum $p_1$. At least three more parameters are needed to describe the emission of interstellar dust along the line of sight in our galaxy and in the cluster. This can be minimally modeled as a thermal spectrum with optical depth $\tau_D$, temperature $T_D$, and  spectral index of emissivity $\beta$. In addition, free-free and synchrotron emission from the diffuse medium in our Galaxy and in the galaxies in the cluster must be accounted for, and one needs at least one amplitude parameter for each of them ($\Delta I_{ff}$, $\Delta I_{sy}$).

Given this situation, it is evident that photometry cannot be sufficient to constrain all these independent parameters, and even to separate the contribution from the thermal SZE from the others. The approach of current experiments is to use etherogeneous ancillary measurements (X-ray surveys, far-IR surveys, optical surveys) to estimate some of the parameters. A direct measurement would certainly be less prone to systematic effects.

As discussed in de Bernardis et al. (\cite{debe12}), low-resolution spectroscopy of the SZE can improve the situation significantly, providing a sufficient number of degrees of freedom to fit several of the parameters above. The price to pay for spectroscopic measurements is that the sensitivity of the measurement is lower than photometric measurements taken in the same conditions.

This is expecially true in the case of FTSs, where radiation from the whole measured frequency range is continuosly detected during the scan of the interferogram. A good trade-off can be obtained in two ways.

First, the instrument should operate from a space-based platform
(stratospheric balloon or satellite) so that only the natural
(astrophysical) radiative background and the emission of the
instrument are present. This is needed anyway, at least at
frequencies outside the mm atmospheric windows and above 250 GHz,
where atmospheric transmission is poor and atmospheric
fluctuations severely limit SZE observations.

Second, the radiative background is further reduced by splitting
the frequency coverage into a few sub-bands, using dichroic filters,
and using one array for each sub-band (see e.g. de Bernardis et al. \cite{debe10}). In
this way, the high background at the high end of the explored
frequency range does not limit the performance of the detector
array measuring the low end. In Fig. \ref{SZE} we plot a sample
splitting of the 0$-$20 cm$^{-1}$ range into four sub-bands
optimized for observations of the SZE with reduced radiative background at balloon altitude.
Wider bands with an increased coverage can be used for space missions, where there is no atmospheric
background. In that case, a cryogenic implementation of the instrument
is needed to fully exploit the reduced natural (astrophysical) background.
Ground-based operation at the best sites are restricted to the 90GHz and 140GHz windows, because of the high radiative background.

In Table \ref{background} we report the typical radiative background due to the
room-temperature emissivity of mirrors (Bock et al.  \cite{Bock95}) and
wire-grids in the spectrometer (Schillaci et al. \cite{schi13}), to the natural
astrophysical background, and to residual atmosphere
for a room-temperature implementation of the instrument
on a good ground-based site and on-board of a balloon payload, and a cryogenic implementation on-board of a satellite.

\begin{table}[h!]
  \caption{Radiative background and photon noise from instrument, atmosphere, and astrophysical background in several sub-bands suitable for SZE measurements. Photon noise per spectral bin is computed assuming that noise from all the frequencies within the sub-band is contributing to the measurement of the considered bin. Top: room-temperature spectrometer on a ground-based telescope (1mm PWV). Middle: room-temperature spectrometer on a stratospheric balloon. Bottom: cold (2K) spectrometer on a satellite. All instruments are diffraction limited at the lowest end of each sub-band. Turbulence and atmospheric fluctuations, which can be relevant in the groud-based case, are not included.}
  \begin{center}
    \begin{tabular}{ccc}
    \hline
& ground & \\
\hline
band &  background   & noise in 1s \\
    (GHz) &  (pW)  & (fW/GHz) \\
    \hline
    85$-$110  & 40 & 55 \\
    125$-$175  & 110 & 110 \\
\hline
    \hline
& stratospheric balloon & \\
\hline
band &  background   & noise in 1s \\
    (GHz) &  (pW)  & (fW/GHz) \\
    \hline
    125$-$175  & 80 & 90 \\
    190$-$260  & 110 & 130 \\
    280$-$360  & 110 & 160 \\
    390$-$420  & 40 & 100 \\
    \hline
\hline
  & satellite & \\
\hline
  band &  background   & noise in 1s \\
    (GHz) &  (pW)  & (fW/GHz) \\
    \hline
    100$-$180  & 7 & 30 \\
    180$-$350  & 10 & 50 \\
    350$-$700  &  16 & 80 \\
    700$-$1000  & 7 & 70 \\
    \hline
\label{background}
    \end{tabular}
  \end{center}
\end{table}

Here we show that combining multiband photometric measurements and spectroscopic measurements obtained by adding a DFTS in front of the same photometer is a very effective approach. In practice, photometric measurements provide high S/N ratio broad-band measurements of the SZE, while spectroscopic measurements provide the required additional degrees of freedom to distinguish the parameters to be measured.

\begin{figure}
\centering
\includegraphics[width=9cm]{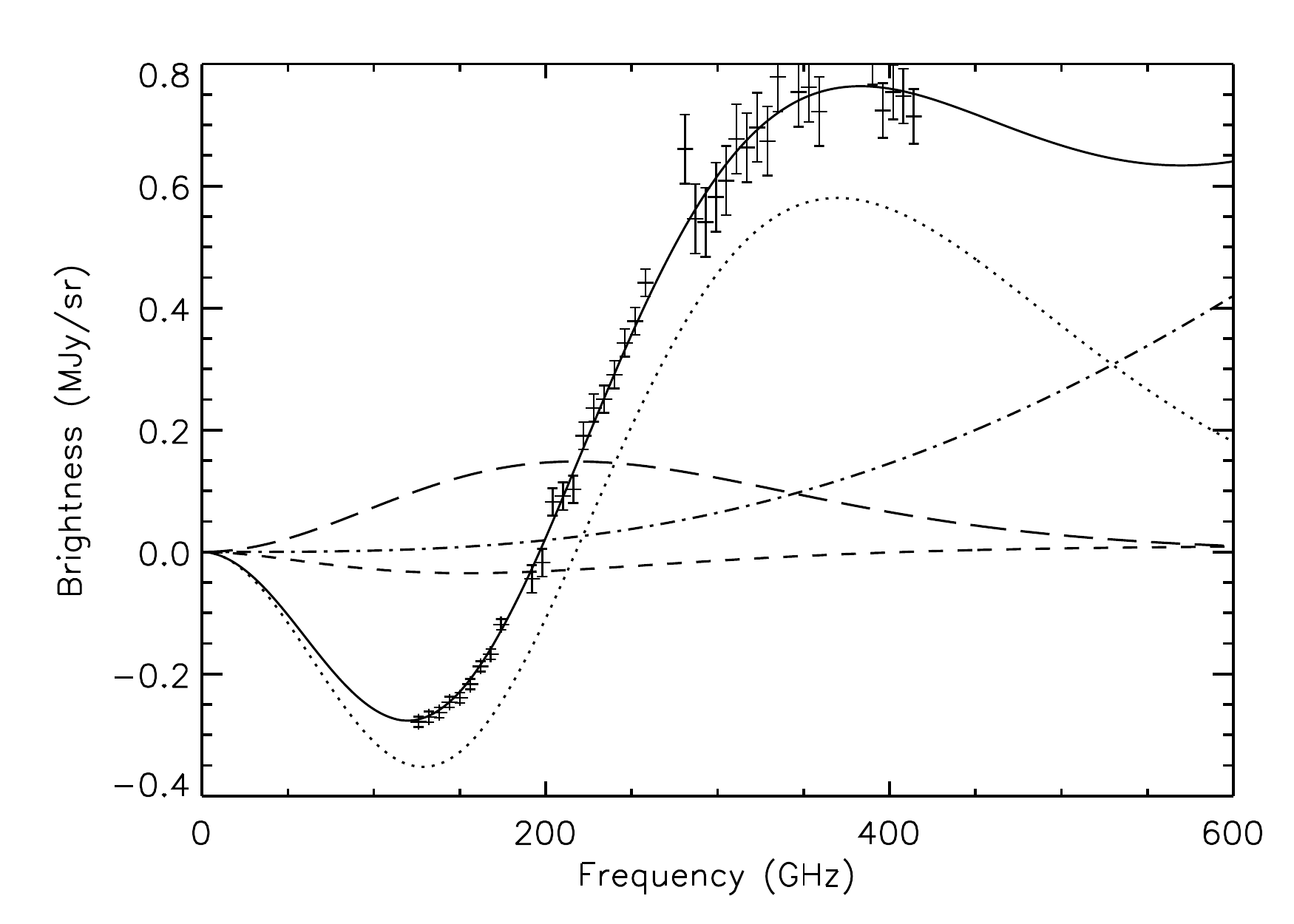}
\includegraphics[width=9.3cm]{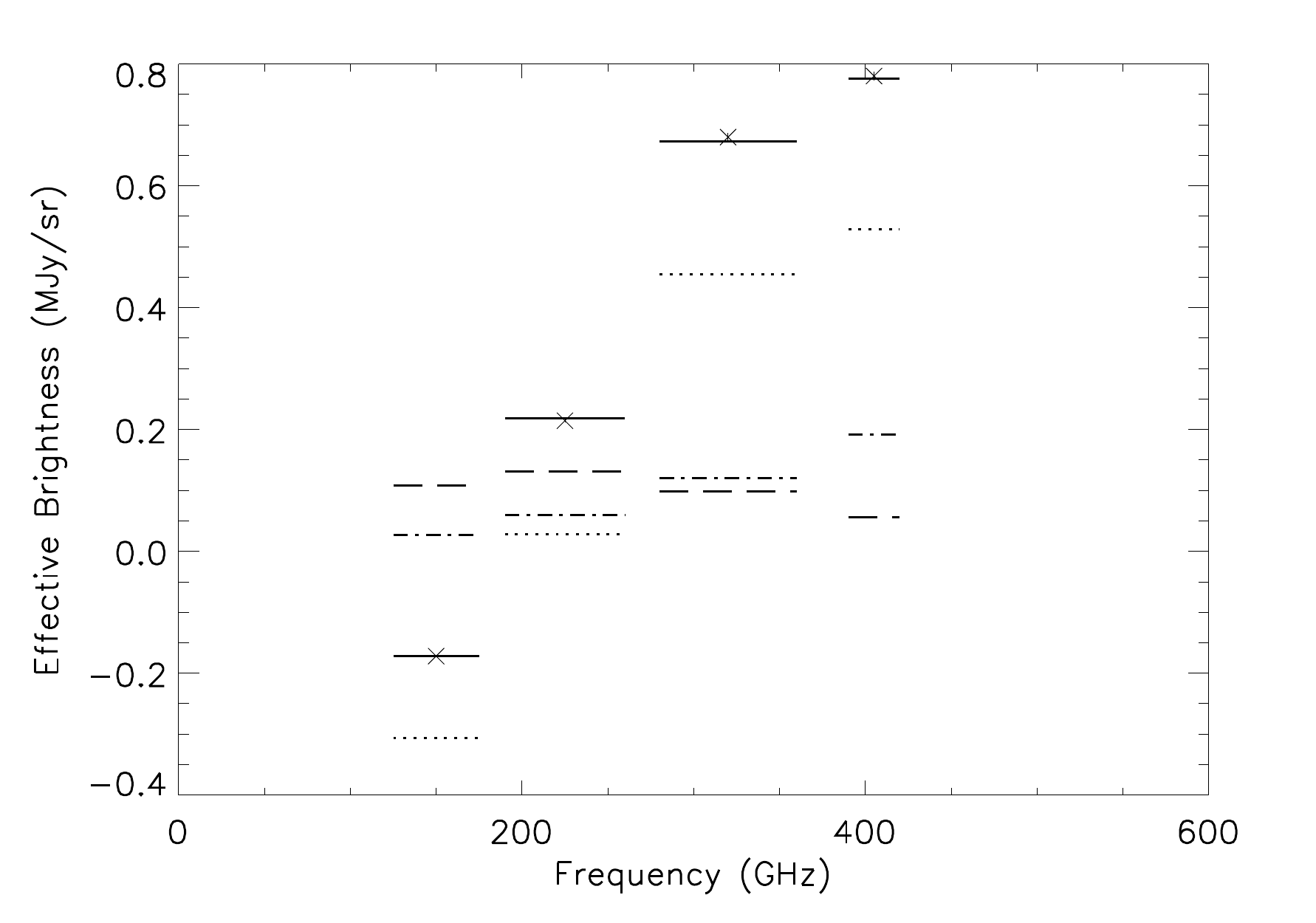}
\caption{Simulated spectroscopic (top) and photometric (bottom) measurements of the SZE towards the center of a typical cluster of galaxies (see Table \ref{fit} for the parameters of the cluster). The lines in the top panel represent simulated spectra of thermal SZE (dotted line),  kinematic SZE + CMB anisotropy (long dashed line), nonthermal SZE (short dashed line), interstellar dust (dot-dashed line), and the best fit to the simulated data using the sum of all the components (continuous line). The data points with their error bars are computed from the sample observation described in the text for a single bolometer of the array observing the center of the cluster. The integration time for spectroscopic measurements is 4 hours.  In the lower panel, simulated photometric measurements are shown, represented by x symbols. The (very small) error bars represent estimates of the 1-$\sigma$ error for an integration time of 4 hours. The bandwidth-integrated power detected in these measurements has been divided by the bandwidth and by the throughput to obtain an effective brightness. The detectors are diffraction limited at the low-frequency edge of the band. Different lines refer to the contribution to effective brightness from different sources (same line-styles as above). The horizontal widths refer to the FWHM photometric bandwidth.   }\label{S+P}
\end{figure}

To investigate this, we reproduced the analysis in de Bernardis et al. (\cite{debe12}) assuming to perform spectroscopic measurements for half of the observation time, and photometric measurements in the same bands for the rest of the observation time, assuming that in both cases the measurements are limited by the noise of the background photons. We also fitted the simulated data with the sum of emission from all the sources along the line of sight, described by the parameters listed above. To estimate the best fit, we minimized the $\chi^2$ obtained by adding the $\chi^2$ of the spectroscopic measurements and the $\chi^2$ of the photometric measurements. The results of this procedure are summarized in Table \ref{fit} for a four-band photometer and spectrometer with bands similar to those of Fig. \ref{SZE}. As a photometer alone, the instrument cannot measure more than four parameters (or independent  combinations). In Table \ref{fit} we demonstrate the power of using the instrument as a spectrometer and as a spectrometer and photometer: five or six parameters can be measured, with improved performance for combined measurements. This demonstrates the synergy of photometric and spectroscopic observations. As described below, our instrument is implementd so that both spectroscopic and photometric observations can be performed in sequence. In Fig. \ref{S+P} we compare data and best fit for the observation simulated in Table \ref{fit}, which corresponds to four hours of integration on a typical cluster with the OLIMPO experiment.

\begin{table}[h!]
  \caption{Cluster and foreground parameters estimated from simulated observations of the SZE along a line of sight towards a cluster of galaxies, using the same bands as in Fig. \ref{SZE}.
S refers to spectroscopic observations (4 hours), S+P refers to spectroscopic observations for 4 hours, and photometric observations for 4 additional hours.  }
  \begin{center}
    \begin{tabular}{lccc}
    \hline
    \hline
   5 parameters & input & best fit (S) & best fit (S+P) \\
    \hline
    $100\tau_T$ &  1.70  & 1.76$\pm$0.32 & 1.76$\pm$0.29 \\
    kT$_e$(keV) &  9.5 & 9.5$\pm$1.7 & 9.6$\pm$1.7 \\
    $10^6 \tau_D$ & 1.85 & 1.85$\pm$0.22 & 1.85$\pm$0.17 \\
    $10^4\Delta T_{CMB}$ & 3.1 & 3.04$\pm$1.26 & 3.07$\pm$0.11 \\
    $10^4 \tau_{nt}$ & 1.0 & 0.95$\pm$2.13 & 1.00 $\pm$0.11 \\
    $\chi^2/DOF$ & -  & 35.8/36  &  39.4/40 \\
\hline
\hline
   6 parameters & input & best fit (S) & best fit (S+P) \\
    \hline
    $100\tau_T$ &  1.70  & 1.78$\pm$0.42 & 1.74$\pm$0.29 \\
    kT$_e$(keV) &  9.5 & 9.6$\pm$1.8 & 9.6$\pm$1.6 \\
    $10^6 \tau_D$ & 1.85 & 1.94$\pm$0.13 & 1.82$\pm$0.23 \\
    $10^4\Delta T_{CMB}$ & 3.1 & 3.07$\pm$1.54 & 3.05$\pm$0.11 \\
    $10^4 \tau_{nt}$ & 1.0 & 0.38$\pm$6.07 & 0.88 $\pm$0.86 \\
    $p$ & 1.0 & 5 $\pm$26 & 3.2 $\pm$6.1 \\
    $\chi^2/DOF$ & -  & 35.8/35  &  42.1/39 \\
\hline
\hline
    \end{tabular}
 \end{center} \label{fit}
\end{table}

\section{Efficient differential MPI} \label{efficient}

\begin{figure}[ht]
\centering
\includegraphics[width=7.5cm]{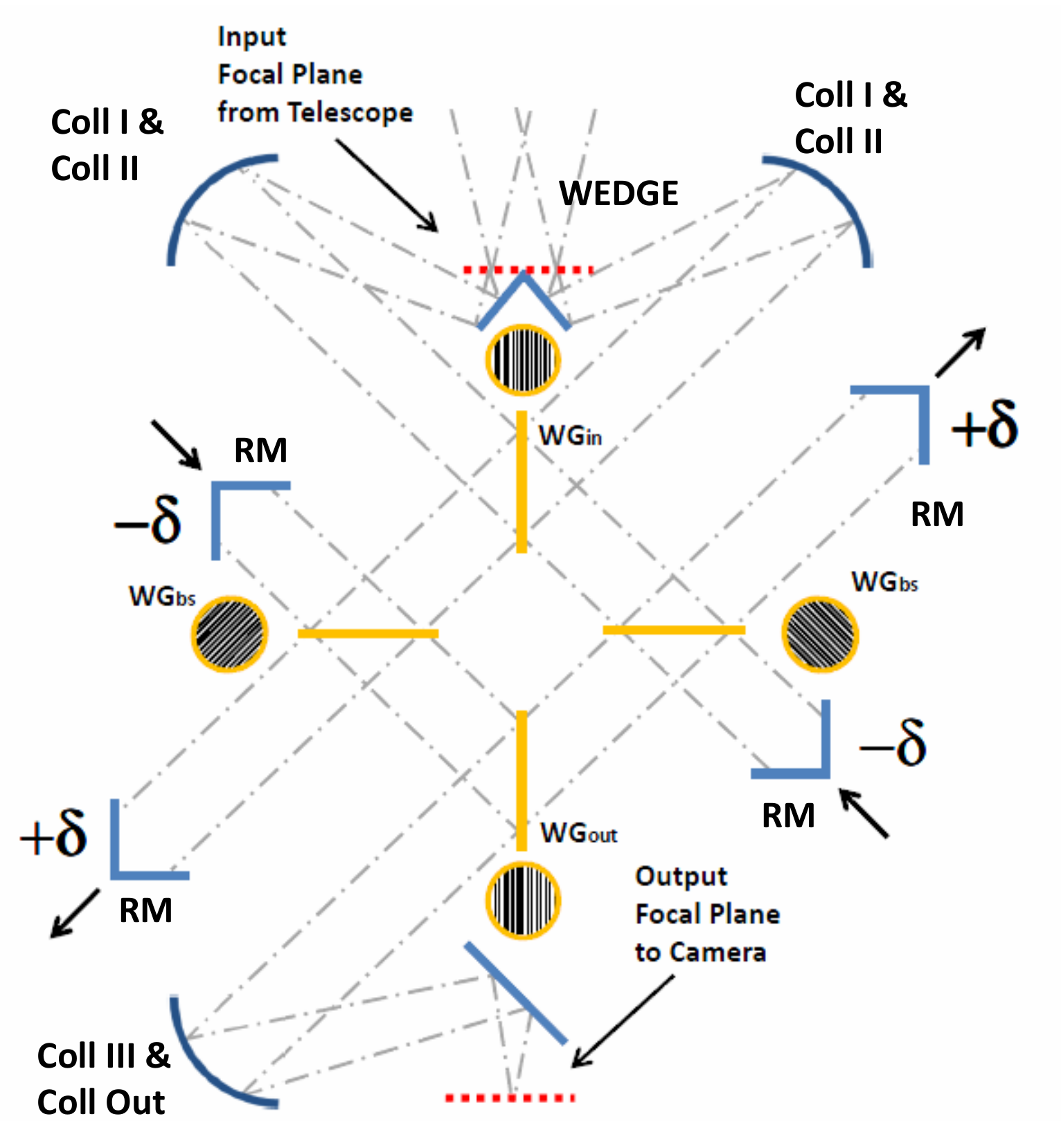}
\caption{Block diagram of the efficient differential MPI.
}\label{schema}
\end{figure}

In practice (see Fig. \ref{schema}), the first component of our spectrometer is a
wedge mirror placed in the focal plane of the telescope, which has
an alt-az mount. The wedge is aligned with the elevation direction.
In this way, radiation from the left half of the focal plane is
reflected into input port 1 of the DFTS, and radiation from the
right half of the focal plane is reflected into input port 2 of
the DFTS. This effectively divides the telescope field of view in
two. For SZE observations one of the two
half-fields is centered on the galaxy cluster, and the other is
centered on a neighboring reference region.

After the constant common mode signal is removed from the interferogram, each detector in the
focal plane of the spectrometer measures the brightness {\sl difference}
between two independent sky pixels symmetrically placed with
respect to the elevation direction through the boresight.
The center of the target field is compared with the center of the
reference field; a pixel east of the center of the target field is
compared with a pixel west of the center of the reference field; a
pixel west of the center of the target field is compared with a
pixel east of the center of the reference field, and so on.

All our implementations of the spectrometer have the same optical
configuration as sketched in Fig. \ref{schema}: a wedge mirror splits
the telescope focal plane in two, as described above, and
two collimators redirect the two beams on opposite sides of the
input polarizer WG $_{in}$. This device is the input polarizer for two MPIs,
one for the radiation reflected and one for the radiation
transmitted. The two MPIs also share the output polarizer WG  $_{out}$ and the
detector arrays, while the beamsplitter polarizers (WG $_{bs,L}$ and WG $_{bs,R}$) and the delay
lines ( $\delta$ ) with the roof mirrors are replicated. Metallic wire grid polarizers are used,
which provide excellent efficiency over the wide frequency range of interest.

We demonstrate that our optical configuration is efficient with the Jones calculus.
In this formalism and following the sketch in Fig. \ref{schema} we can
write the electrical fields for the two orthogonal polarizations coming from the two sources A and B as
$$ A=\left(
    \begin{array}{c}
      A_x \\
      A_y \\
    \end{array}
  \right); \ \
  B=\left(
    \begin{array}{c}
      B_x \\
      B_y \\
    \end{array}
  \right) \ .
$$

Here the reference system is right-handed and comoving, following the propagation of the electromagentic wave, with the $z$ axis antiparallel to the wavevector ${\vec k}$, and the $x$ axis in the plane of the drawing in Fig. \ref{schema}.

The matrix describing a polarizer is :

$$
T_\theta=\left(
  \begin{array}{cc}
    \cos^2(\theta) & \cos(\theta)\sin(\theta) \\
    \cos(\theta)\sin(\theta) & \sin^2(\theta) \\
  \end{array}
\right) $$

when the polarizer transmits radiation and

$$
R_\theta=\left(
  \begin{array}{cc}
    \sin^2(\theta) & -\cos(\theta)\sin(\theta) \\
    \cos(\theta)\sin(\theta) & -\cos^2(\theta) \\
  \end{array}
\right)
$$

when it reflects radiation.

Here $\theta$ is the angle between the principal axis of the polarizer and the $y$ direction (orthogonal to the plane of the drawing in Fig. \ref{schema}, and projected in the $xy$ plane). The matrix for the ideal roof-mirror (sequencing two metal reflections) is the Identity Matrix ${\cal I}$.

The phase shift $\delta$ introduced by the optical path difference and common to both polarizations is described by the matrix

$$
\Delta_{\delta}=\left(
  \begin{array}{cc}
    e^{i\delta} & 0 \\
    0 & e^{i\delta} \\
  \end{array}
\right) \ ,
$$

where $\delta$ is $4\pi x \sigma$, with $x$ the
mechanical path difference in the case of a single moving mirror, and $\sigma$ the wavenumber (in $cm^{-1}$).
In our implementation of the spectrometer, both delay lines have moving mirrors (moving in opposite directions), so we gain a factor two in the introduced phase-shift for a given mechanical displacement of each mirror.

Following Fig. \ref{schema}, radiation from source $A$ is processed by the left FTS after being transmitted by the input wire grid. The processed wave at the left detector array can be written as the sum of the contributions of two beams, traveling in the two arms of the left FTS,

\begin{eqnarray} \label{EIA}
E_{IA}=R_0  T_{45}   \Delta_{\delta}   {\cal I}
R_{-45}   T_0   A
+  R_0   R_{45}
\Delta_{-\delta}   {\cal I }   T_{-45}   T_0   A \ .
\end{eqnarray}

Analogously we find, for the radiation of source B processed by the left FTS

\begin{eqnarray} \label{EIB}
E_{IB}=R_0   T_{45}   \Delta_{\delta}   {\cal I}
R_{-45}   R_0   B
+  R_0   R_{45}
\Delta_{-\delta}   {\cal I}   T_{-45}   R_0   B \ .
\end{eqnarray}

After simple algebra we derive:

 $$ \left(
  \begin{array}{c}
    E_{Ix} \\
    E_{Iy} \\
  \end{array}
\right)
=
\left(
  \begin{array}{c}
    0 \\
    B_y \cos(\delta) + i A_x \sin(\delta) \\
  \end{array}
\right) \ . $$

For the right FTS, equations (\ref{EIA}) and (\ref{EIB}) become
\begin{eqnarray}\label{EIIA}
E_{IIA}=T_0   T_{-45}   \Delta_{-\delta}   {\cal I}
R_{45}   R_0   A
+ T_0   R_{-45}
\Delta_{\delta}   {\cal I}   T_{45}   R_0   A
\end{eqnarray}
and
\begin{eqnarray}\label{EIIB}
E_{IIB}=T_0   T_{-45}   \Delta_{-\delta}   {\cal I}
R_{45}   T_0   B
+ T_0   R_{-45}
\Delta_{\delta}   {\cal I}   T_{45}   T_0   B \ .
\end{eqnarray}

Likewise, for the right FTS,

$$ \left(
  \begin{array}{c}
    E_{IIx} \\
    E_{IIy} \\
  \end{array}
\right)
=
\left(
  \begin{array}{c}
    B_x \cos(\delta) + i A_y \sin(\delta) \\
    0 \\
  \end{array}
\right) \ . $$

The output signal detected by the left-side detector is the sum of the signals from the two MPIs:

$$
\left(
  \begin{array}{c}
    E_{x} \\
    E_{y} \\
  \end{array}
\right)
=
\left(
  \begin{array}{c}
    B_x \cos(\delta) + i A_y \sin(\delta) \\
    B_y \cos(\delta) + i A_x \sin(\delta) \\
  \end{array}
\right) \ . $$
This means for the intensity
\begin{equation}
I_L = |E_{x}|^2+|E_{y}|^2
\end{equation}

\begin{equation}\label{5}
I_L = \frac{1}{2}(I_a+I_b)+\frac{1}{2}(I_a-I_b)\cos(\delta)  \ ,
\end{equation}

where the common mode and the modulated terms in the interferogram can easily be recognized. This is twice the intensity detected in a standard MPI (see e.g. Carli and Mencaraglia \cite{Carl81a, Carl81b}).  A complementary expression is found for the right-side detector array:

\begin{equation}\label{5}
I_R = \frac{1}{2}(I_a+I_b)-\frac{1}{2}(I_a-I_b)\cos(\delta) \ .
\end{equation}

\section{Optical design} \label{design}

Observations of clusters of galaxies require an angular resolution $\theta \sim 1^\prime$.
At the center frequency of the range of interest here (350 GHz) the size of the entrance pupil of the telescope
should be $ D \simeq 1.22 \lambda / \theta \simeq 3.5m$.
A detector array should be placed in the focal plane of the telescope for fast mapping,
covering the entire field of view of the instrument.
This should be $\Theta \sim 15^\prime$ wide to obtain complete images of the cluster and
the surrounding reference areas.
To limit the physical size of the detector array, which has to be cooled inside
a cryostat for high sensitivity, the output focal ratio $f_{\#}$ of the telescope should not be too slow.
Assuming a diameter of the detector array $d \sim 50 mm$ we derive a focal length $F = d/\theta \simeq 11.5 m$,
so that $f_{\#}=F/D\sim 3.3$. This will be the entrance $f_{\#}$ for our spectrometer,
which is to be inserted between the focal plane of the telescope and the detectors array.

\begin{figure}[ht]
\centering
\includegraphics[width=7.5cm]{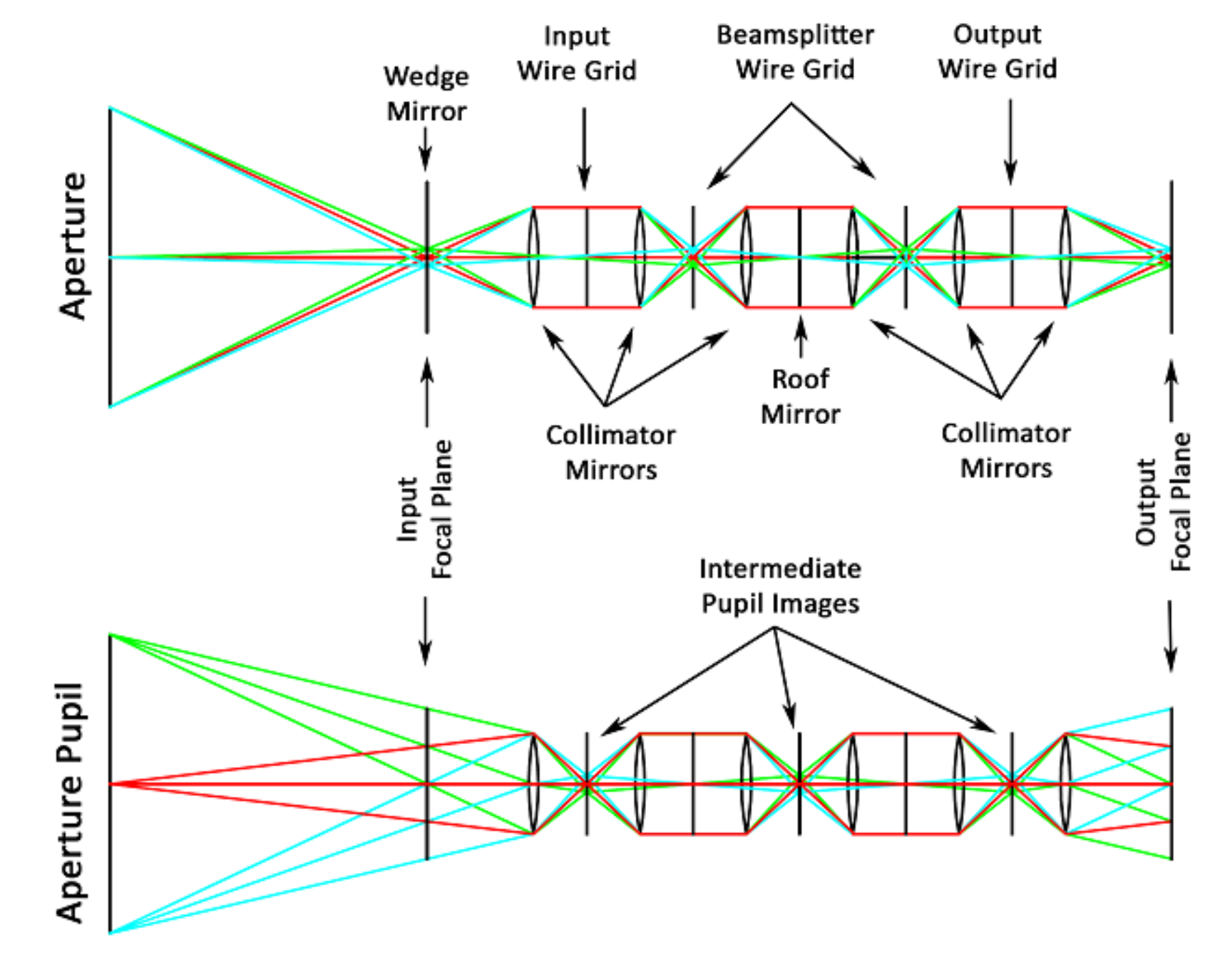}
\caption{Propagation of rays in the spectrometer. Top: rays coming from the sky. Bottom: rays coming from the aperture pupil.
}\label{optics_concept}
\end{figure}

In Fig. \ref{optics_concept} we show the path of the radiation inside the
spectrometer from two points of view: in the upper diagram we show the
path of rays coming from the sky, while in the bottom diagram we show the path of
rays coming from the aperture pupil. In this sketch collimators are represented by lenses
for simplicity.

Following the upper diagram, the feed optics (the
telescope) focuses the image of the sky at the entrance of the FTS.
This, in turn, produces  an exit beam with the same $f_{\#}$.
In this way, the FTS can be removed to return to
photometric measurements.

In the spectrometer we need to control the divergency of the beam,
which is directly related to the dimensions of the optical components.

Moreover, when we move the roof mirrors to change the optical path difference,
the optical configuration can change. To mitigate this problem we need to keep
the beam collimated at the level of the roof mirrors. For this reason we need
two intermediate focalizations in correspondence to the two
beam-splitting wire grids (WG$_{BS}$).

In the bottom diagram of Fig. \ref{optics_concept} we show the positions of the
intermediate images of the aperture pupil (the telescope aperture): these
correspond to positions where the rays from the sky are collimated.
This optical scheme requires three intermediate images: two at the
WG $_{in}$ and WG  $_{out}$ and a third one at the movable roof mirror.
In this way, the effect of optical pickup or
modulation is minimized when moving the mirror.

The considerations above have driven the optical design of the instrument.

\section{Opto-mechanical implementation} \label{mech}

\subsection{Optics optimization} \label{optimization}

The initial optical design of the efficient DFTS was developed using ZEMAX in \emph{sequential mode}. We modeled the optical system that feeds the spectrometer as a classical Cassengrain telescope, feeding a focal plane $50mm$ in diameter, with a focal ratio $f/3.3$. We need to insert the input ports of our DFTS in this beam, between the rear side of the primary mirror and the telescope focal plane, and produce an identical beam at the exit, so that the insertion of the DFTS does not require any change in the detector system (see Fig. \ref{figura1.od}). In this way, the telescope can carry out photometric measurements when the DFTS is excluded from the system, and spectroscopic measurements when the DFTS is included. In the following we describe the optimization process for the different sections of the instrument.



\begin{figure}[ht]
\centering
\includegraphics[width=8.5cm]{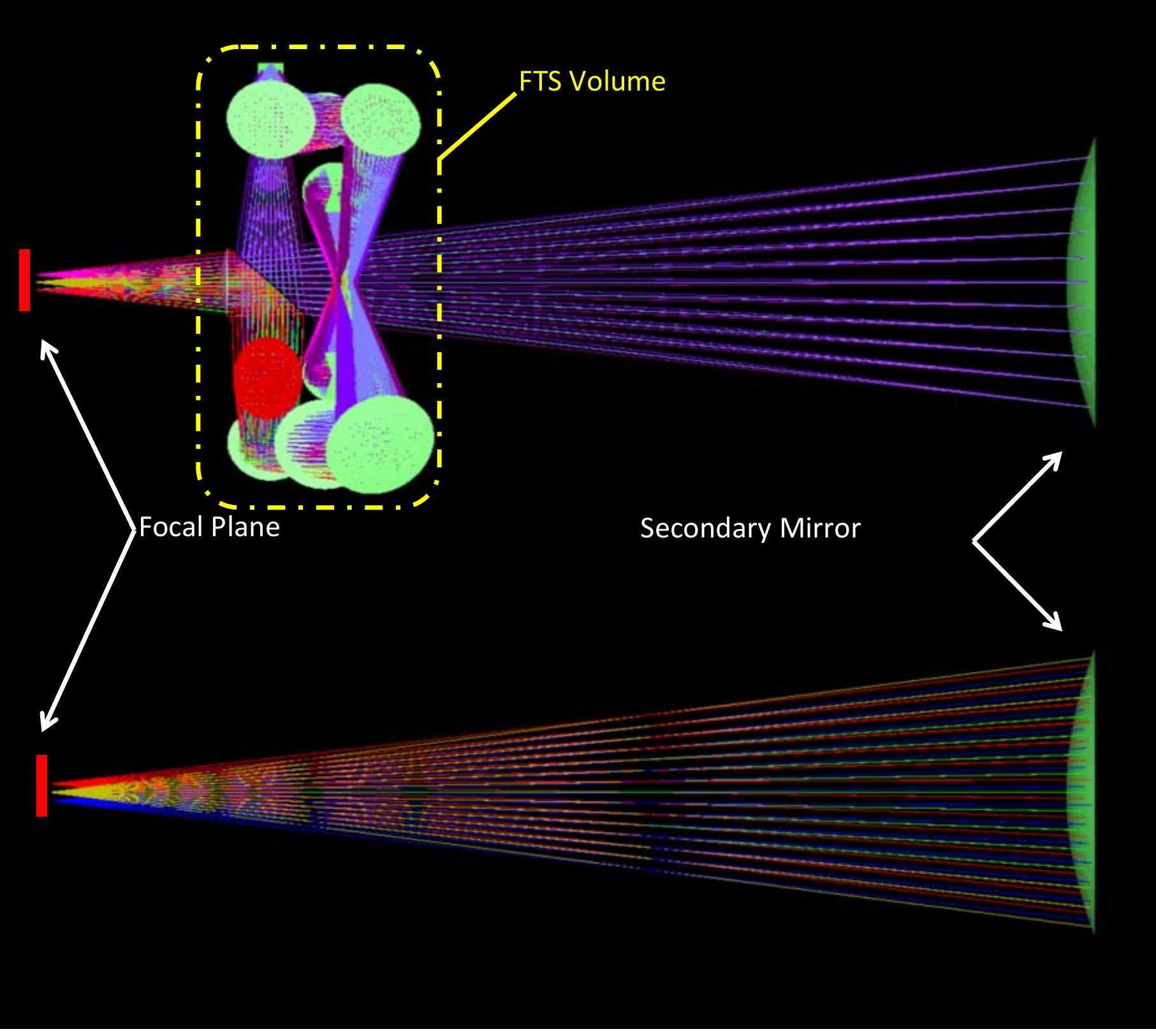}
\includegraphics[width=8.5cm]{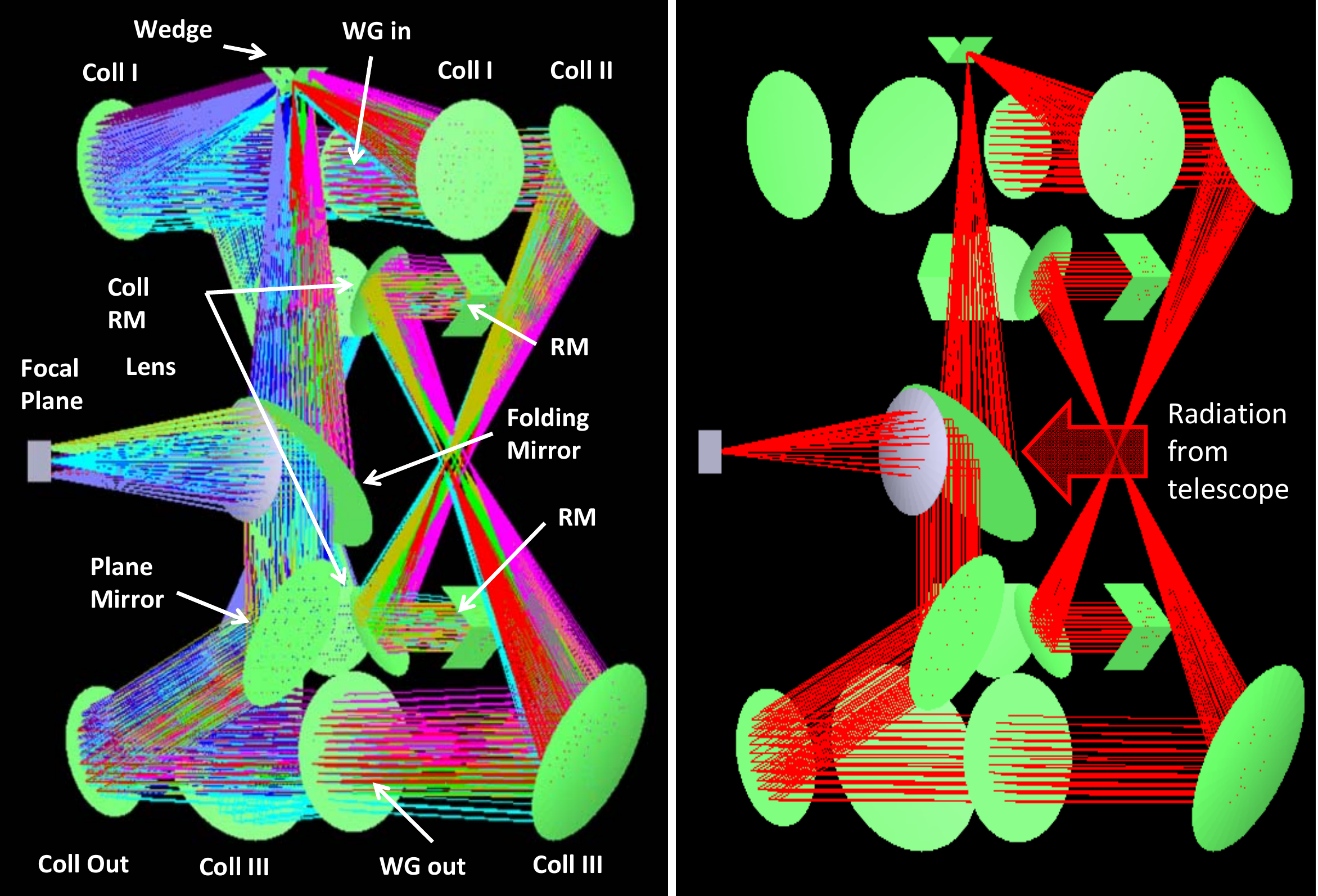}
\caption{{\bf Top}: location of the interferometer between the secondary mirror and the focal plane inside the cryostat (top); ray-tracing with the interferometer excluded (bottom). {\bf Bottom left}: general ray-tracing in the interferometer. {\bf Bottom right}: ray-tracing of a single field in one FTS. Radiation coming from the telescope is reflected by the 45$^o$ folding mirror towards the wedge mirror.  A first collimator (Coll I) reflects it towards the input Wire Grid (WG in) that splits the polarizations towards the two inteferometers. In each FTS radiation from this first WG proceeds towards a second collimator (Coll II) that focuses it on the beam splitter (WG BS). Both the reflected and transmitted components are collimated by collimator mirrors (Coll RM) on the Roof Mirrors (RM) of the delay lines. The roof mirrors rotate the polarization by 90$^o$ so that the beams proceed towards the output section, where Coll III collimates the beam on the output wire grid (WG Out). Radiation is finally collimated by Coll Out and reflected by the rear face of the 45$^o$ folding mirror so that the exit is aligned with the optical axis of the telescope.  A HDPE lens adapts its divergency to the initial one.
}
\end{figure}
\label{figura1.od}

\begin{figure}[ht]
\centering
\includegraphics[width=9cm]{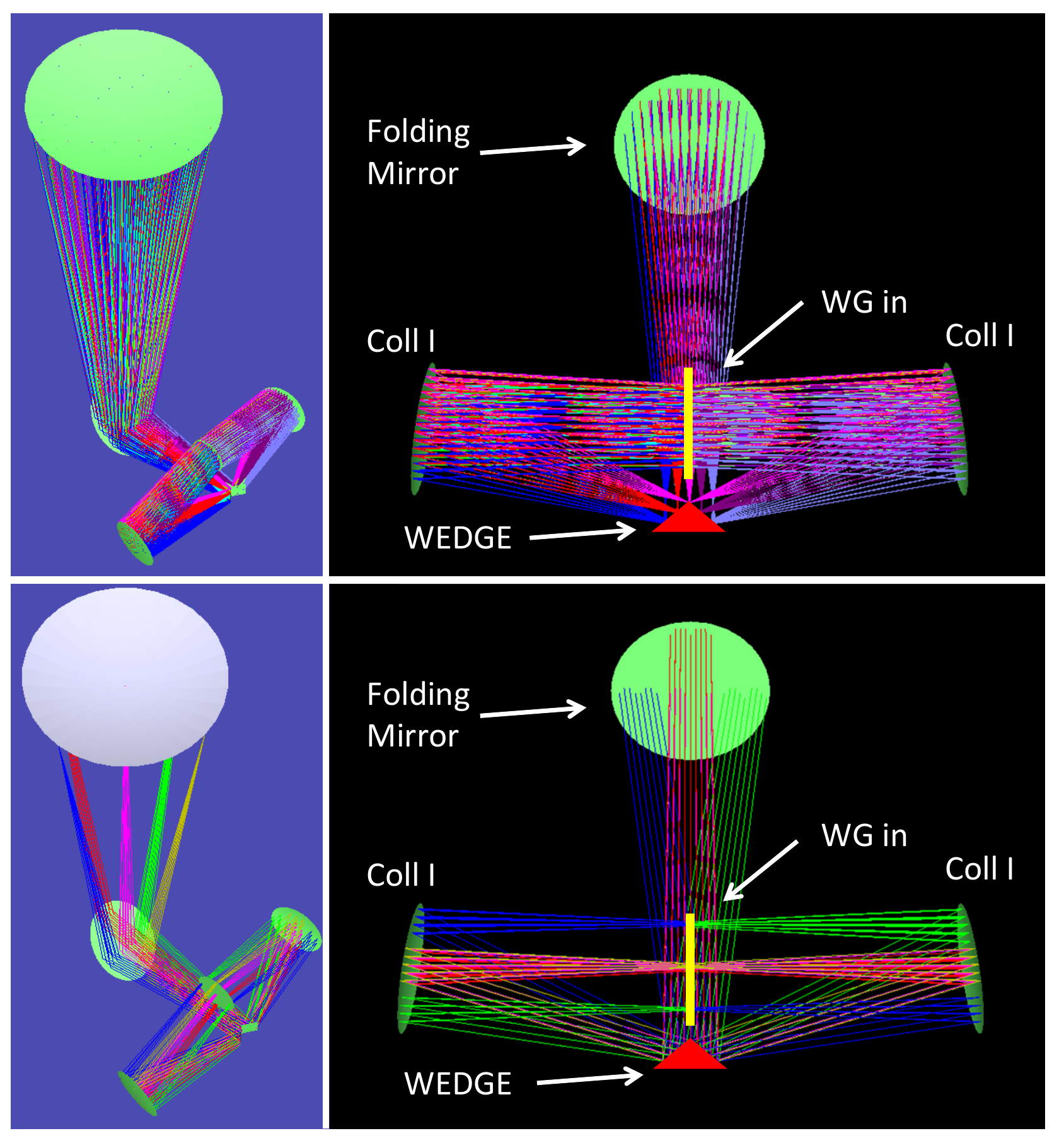}
\caption{Input section of the DFTS. Top: rays from the sky. Bottom: rays from the secondary mirror of the telescope.
}\label{figura2.od}
\end{figure}

\subsection{Input section}

Radiation coming from the Cassengrain secondary mirror is deflected towards the input port of the DFTS by a flat folding-mirror at $45^o$. To better fit the available room between the rear side of the primary mirror of the telescope and the cryostat of the detectors, the optical system of the DFTS is developed in a plane orthogonal to the optical axis of the telescope (see Fig. \ref{figura3.od}).

The input section of the DFTS is shown in Fig. \ref{figura2.od}.
In the location of the deflected focal plane we place our \emph{image divider}. A wedge mirror, with vertex in the focal plane, splits the image of the sky in two (see Fig. \ref{figura2.od}). The two halves are reflected towards two independent collimators (Coll$_{I}$), symmetrically located with respect to the focal plane.

The two sides of the mirror are $40mm \times 40mm$, and they are tilted $\pm 36^o$ with respect to the optical axis. The two Coll$_{I}$  have a \emph{biconic} surface, optimized to reproduce the aperture pupil at the input wire grid WG $_{in}$  position.

Below we consider the center of the field of view of the single half of the wedge as our optical axis. On WG $_{in}$ we have the image of the secondary mirror of the telescope, that is, the aperture stop of the optical system (see Fig.  \ref{optics_concept}). The wire grid diameter required to avoid vignetting of the beam is $130 mm$. The two output beams from WG $_{in}$ are fully polarized; both contain radiation from the two input sources and are processed by two symmetrical MPIs. In the next subsection we follow the radiation path that propagates and interferes inside the two MPIs.

\subsection{Interferometer section}

\begin{figure}[ht]
\centering
\includegraphics[width=8.9cm]{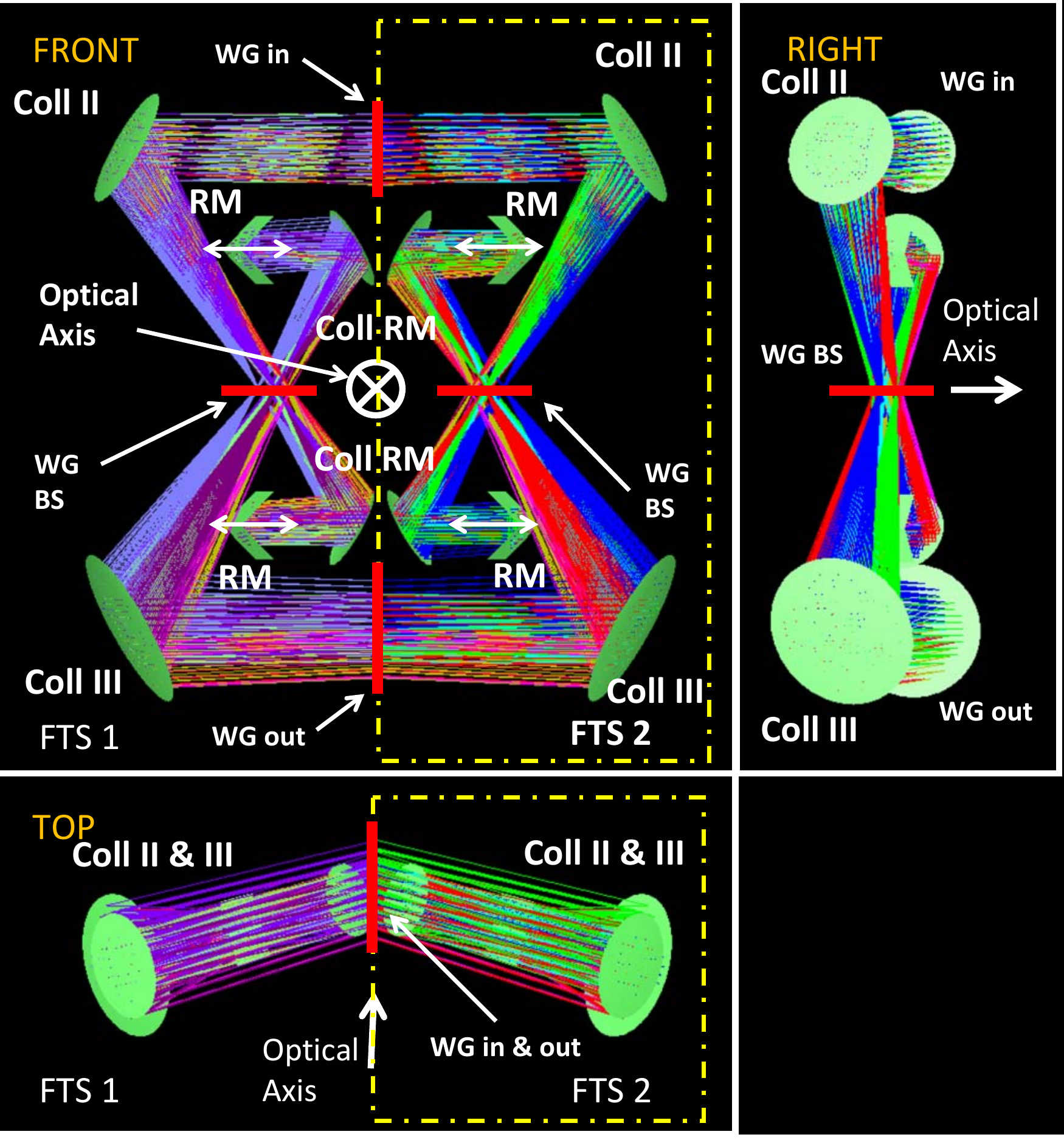}
\caption{Orthogonal views of the two MPIs. The optical axis of the Cassegrain telescope is marked for reference. The moving roof mirrors of the 4 delay lines are marked by double arrows.
}\label{figura3.od}
\end{figure}
The optical design of the MPIs is shown in Fig. \ref{figura3.od}.
Undesired rotation of the polarization orientation is avoided by developing the two MPIs along two slanted planes. These are tilted with respect to a plane normal to the telescope optical axis by $\pm 16^o$.  As evident from Fig. \ref{figura3.od}, we also need to preserve a volume along the telescope optical axis for the folding mirror that intercepts the telescope beam. This limits the available space for the delay lines that have to be folded up to avoid vignetting.

The two MPIs are completely symmetric. The outgoing beam from WG $_{in}$ is collimated and directed towards the beamsplitter wire grid (BS) and the delay lines. Owing to its divergency, we need to use two additional aspheric collimators that minimize the dimensions of the optical system and the aberrations and limit the effect of the retro-reflector motion.
A second biconic collimator (Coll$_{II}$) focuses radiation onto the BS; a parabolic collimator (Coll$_{RM}$) produces the parallel beam to illuminate the two roof mirrors (RM) retro-reflectors of each MPI. The optimization performed allows us to limit the diameter of the BS wire grid to $50mm$. The room available for the delay lines allows a $40mm$ mechanical travel for each RM. The opposite motions of the two RMs in the same MPI double the optical path difference, so that we reach a maximum spectral resolution of 1.9 GHz per spectral element.

In the RMs the polarization axis of the incident beam is rotated by $90^o$, so that, after reflection on the RM, the beam that had been reflected by the BS is sent back to the BS, which will transmit it, while the beam that had been transmitted by the BS will be reflected.
Both beams are directed towards the output polarizer (WG  $_{out}$) by means of a third biconic collimator mirror (Coll$_{III}$).  This has the same optical parameters as Coll$_{II}$, but a wider size to avoid vignetting. Consequently, WG  $_{out}$ has a diameter of $180mm$.
This last wire grid combines the power from the two MPIs. With this configuration the power at the two output ports of the DFTS is close to $100\%$ of the input power.

In the next subsection we describe the output section of the DFTS.

\subsection{Output section}

\begin{figure}[ht]
\centering
\includegraphics[width=7cm]{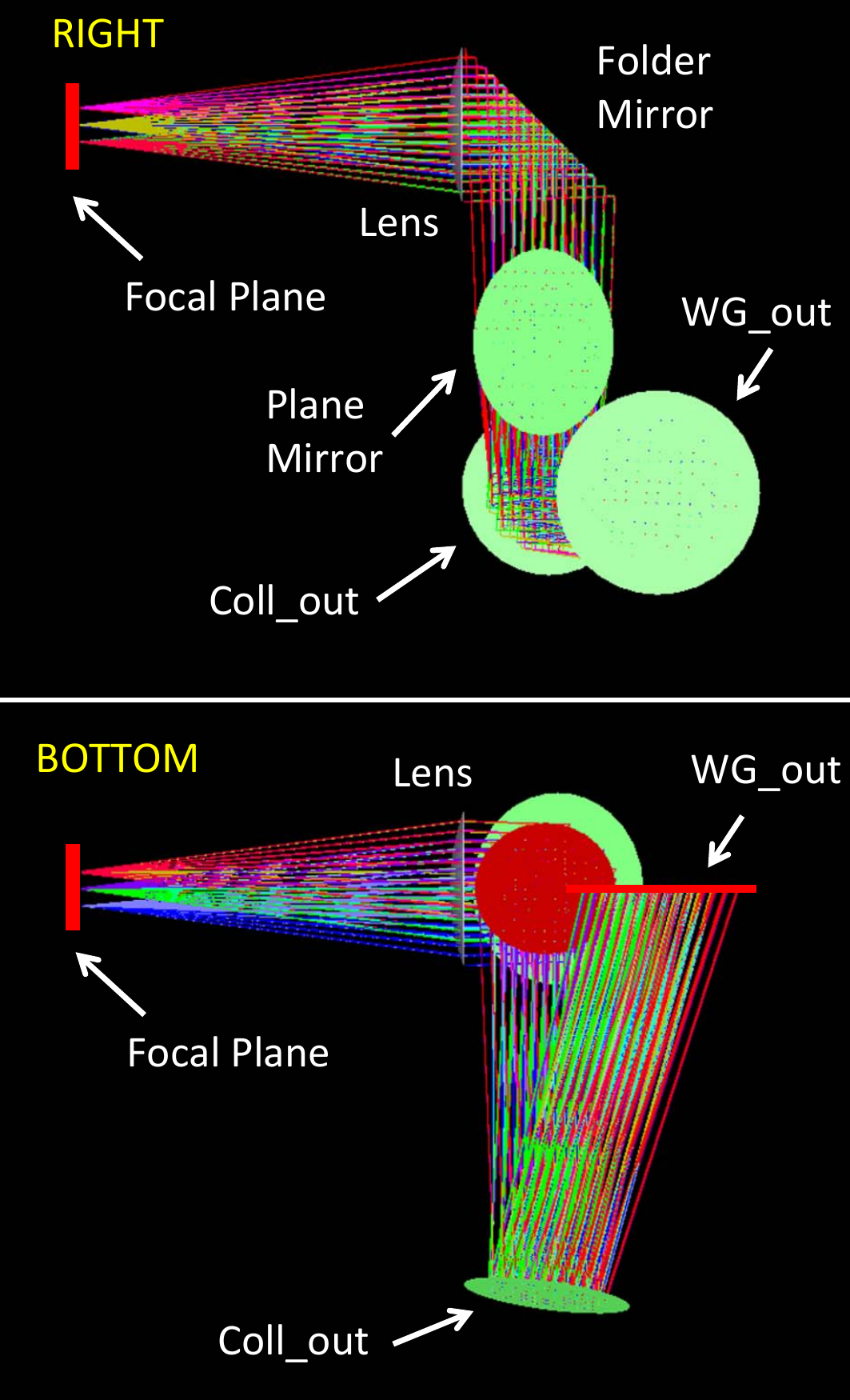}
\caption{Output section of the DFTS}\label{figura4.od}
\end{figure}

The output of WG  $_{out}$ contains power from the two input sources (the two halves of the sky field of view split by the wedge mirror) and from the two MPIs. Each of the two outputs contains $50\%$ of the incident radiation collected by the telescope.

We used only one of the two output beams and focused it to produce the original output focal ratio of the telescope ( $f/3.3$ ). This was obtained by means of an aspherical mirror (Coll  $_{out}$) and a flat mirror, redirecting the beam onto the rear face of the same $45^o$ folding mirror we used to intercept the beam from the telescope (see Fig. \ref{figura4.od}).
Coll  $_{out}$ has a biconic profile with very large radii to optimize the final focalization of the beam. This is completed by a plano-convex lens (made of high-density polyethylene) that corrects aberrations and reproduces the original focal ratio of the telescope.

\subsection{Imaging performance}

\begin{figure}[ht]
\centering
\includegraphics[width=9cm]{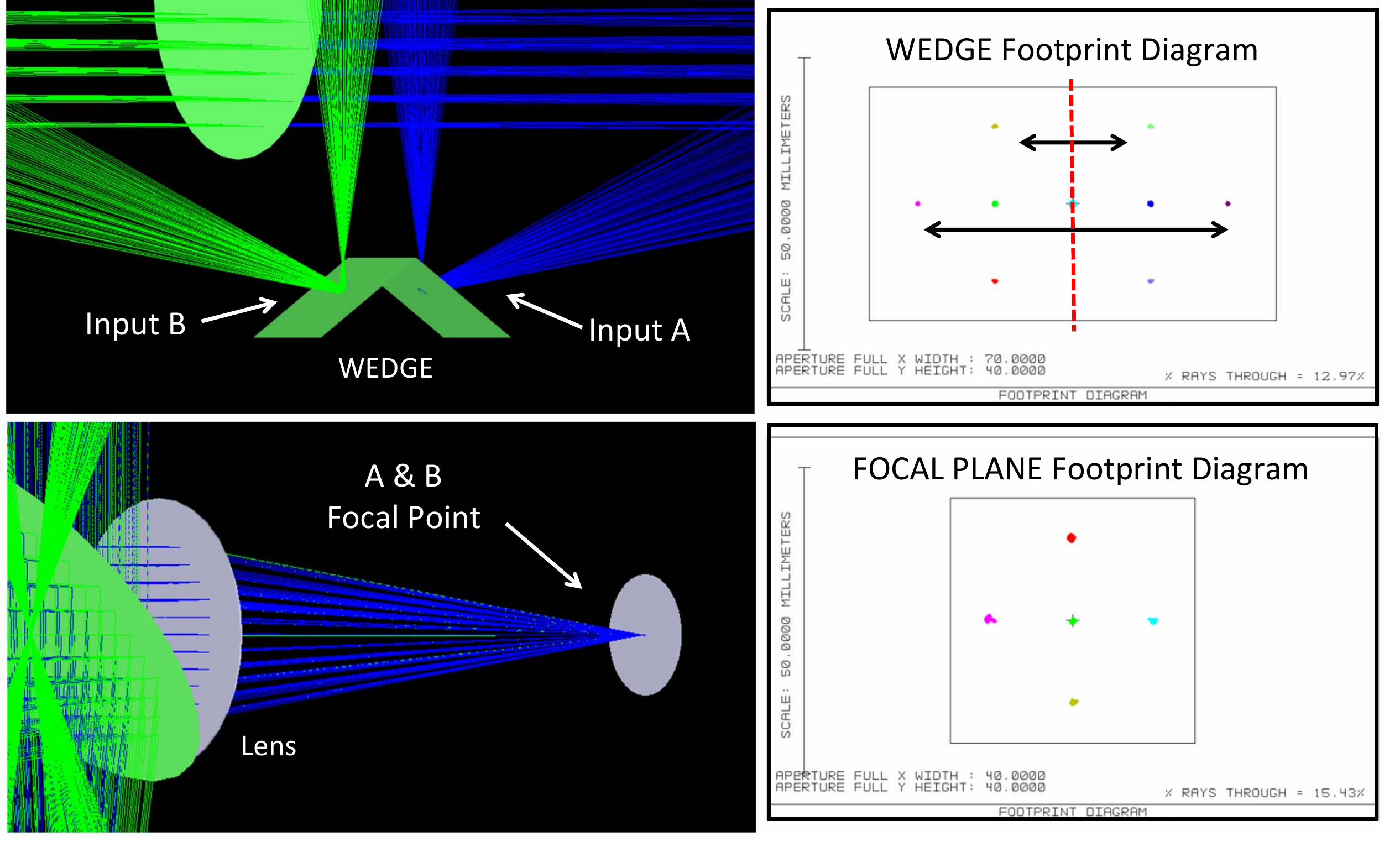}
\caption{Combination of pixel pairs from the right and left half of the telescope focal plane (where the wedge mirror is located, top right diagram) into single pixels in the detector focal plane (bottom right diagram). Images of the ray-tracing near the wedge mirror (top left) and near the detectors focal plane (bottom right) are also shown. }\label{figura5.od}
\end{figure}

The DFTS was designed to have imaging capabilities. The FOV of the instrument is defined by the fraction of the focal plane covered by the wedge mirror at the entrance of the DFTS. In Fig. \ref{figura5.od} (right) we can compare the configuration of the input fields on the wedge mirror level to the configuration of the same fields on the detectors focal plane. The two center ray-traced fields come from the centers of the two faces of the wedge, and, after being processed by the DFTS, are focused on the center of the detector focal plane. The rectangular area represents the surface of the wedge and the colored spots are the field arranged in a pattern that covers a sky area of $0.22^o \times 0.11^o$. The image of the sky is split along the red dotted line, and in the focal plane we have a differential spectral image, where each spot represents the difference between mirrored sky positions.

\begin{figure}[ht]
\centering
\includegraphics[width=8cm]{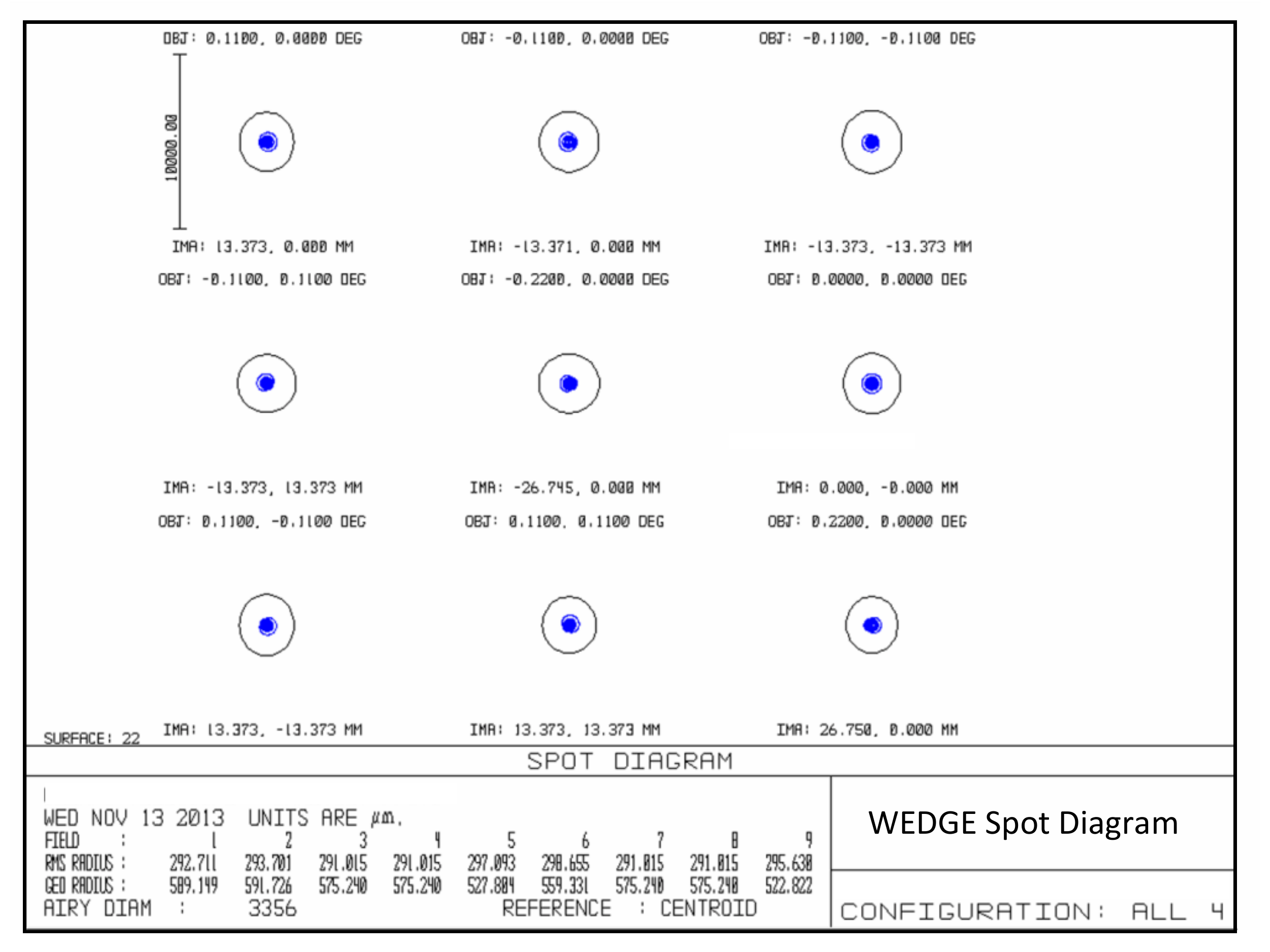}
\includegraphics[width=8cm]{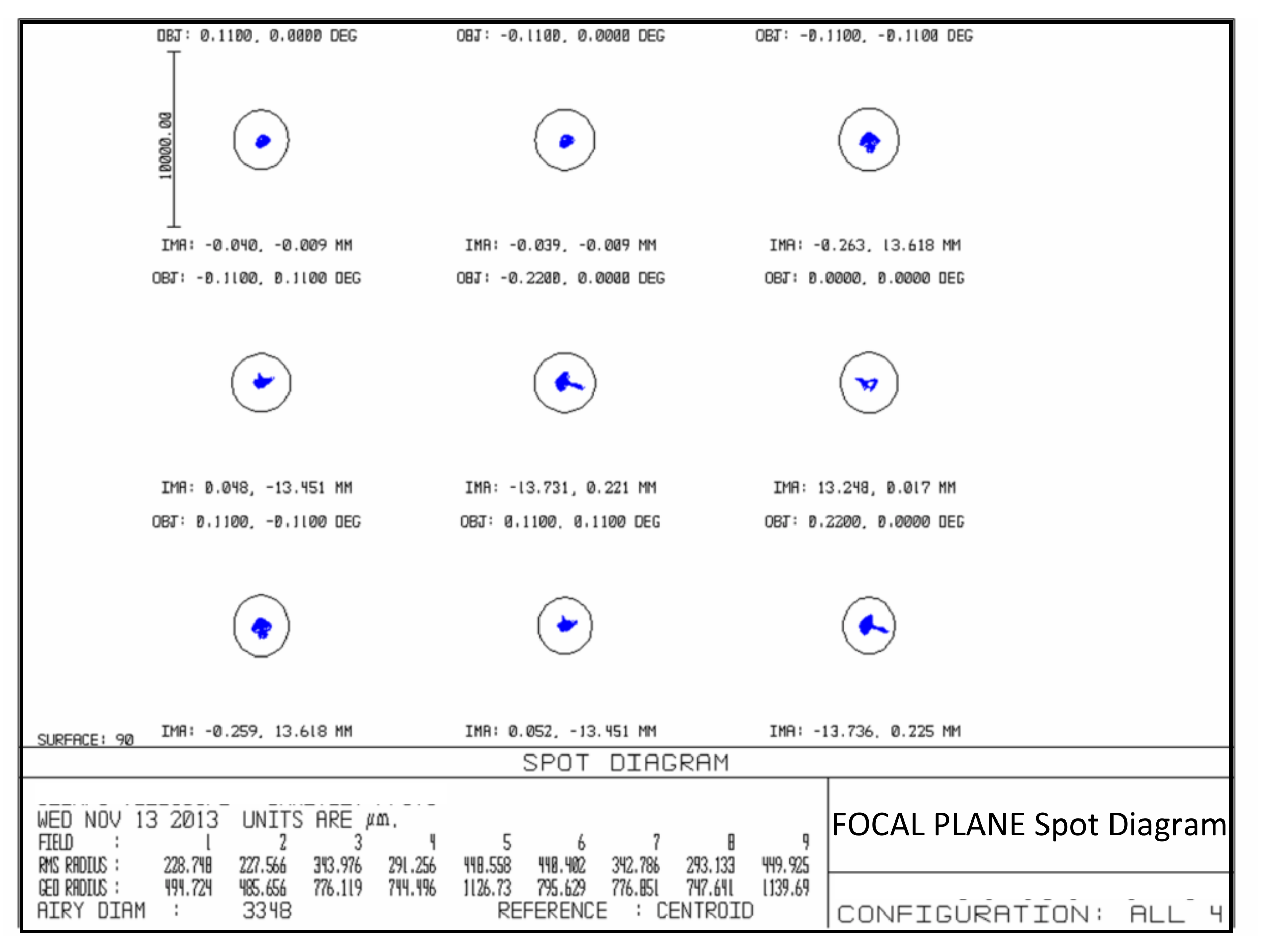}
\caption{Spot diagrams at the input and output of the DFTS. The dispersion of the points is due to the geometrical aberrations resulting from the sum of all the mirrors in the telescope (top panel) and in the DFTS (bottom panel).  The circles represent the Airy disk at a frequency of 600 GHz. The nine fields in each panel refer to positions in the sky covering a rectangular area of 26,4 x 13,2 arcminutes, that is, the FOV of the instrument.  }\label{figura6e7.od}
\end{figure}

Our optical optimization preserves the initial optical quality, as evident from the spot diagrams in Fig. \ref{figura6e7.od}: both at the wedge mirror and at the final focal plane all the fields spots are well inside the Airy disk, for a frequency up of $600 GHz$.

\subsection{Plug-in module} \label{plug-in}

Our instrument has been designed as a plug-in optical processing module for the existing OLIMPO instrument. For this reason, the optical elements assembly including the 45$^o$ folding flat mirror and the output lens can be inserted into or removed from the optical path between the telescope and the detector cryostat. When the assembly is inserted, the front surface of the folding mirror reflects the beam coming from the telescope into the spectrometer, and its rear surface reflects the spectrometer-processed beam towards the lens and the detector cryostat. When the assembly is removed, the spectrometer is excluded, and the OLIMPO instrument works as a four-band photometer array.  The folding mirror / lens assembly is mounted on a trolley on rails (SKF model LWRM/V) that is moved orthogonally to the beam by a linear actuator, on command. The mechanical system provides a position repeatibility of a few microns when inserting the folding mirror in the optical path.

\begin{figure}[ht]
\centering
\includegraphics[width=6.3cm, angle=90]{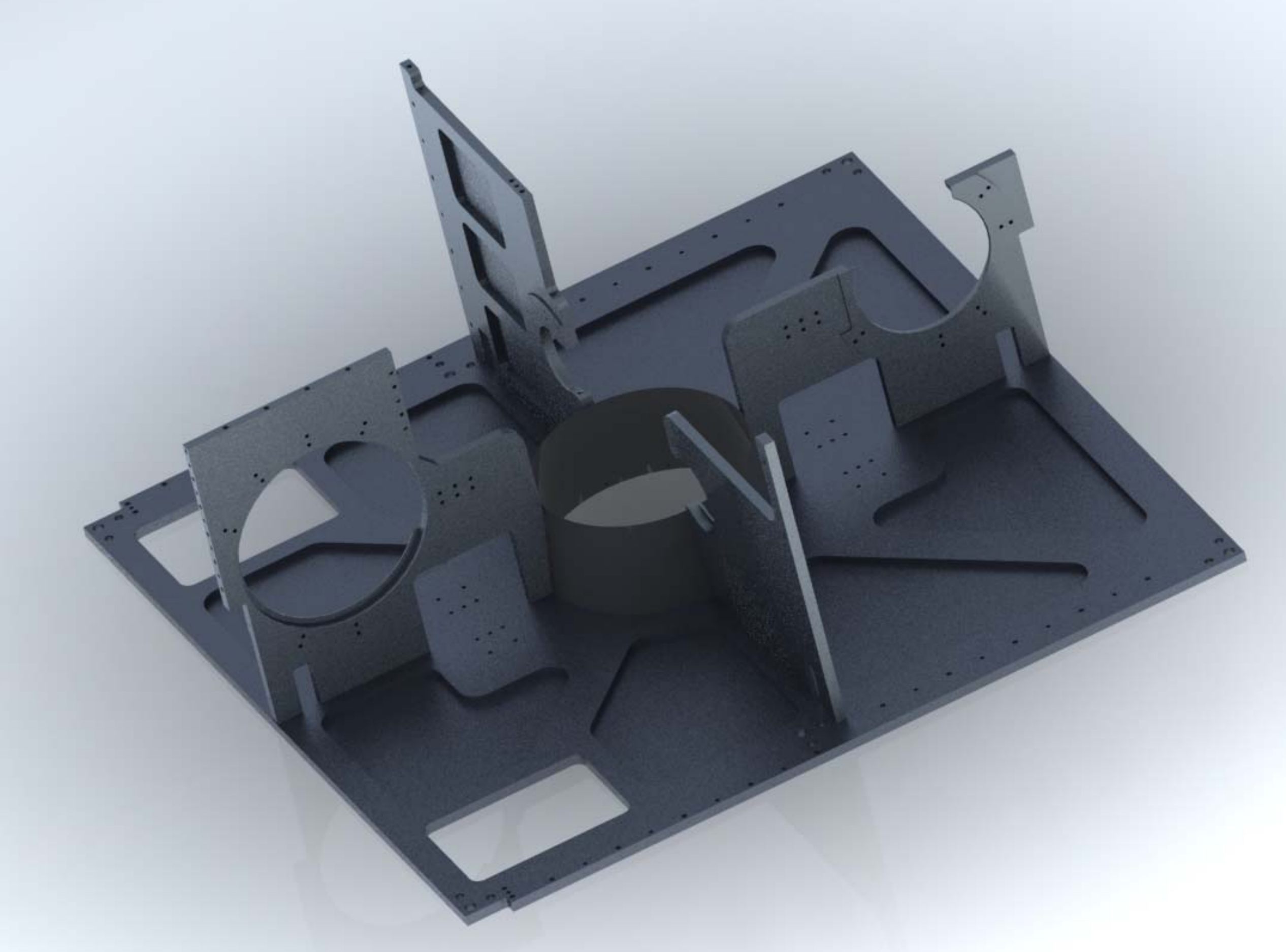}
\includegraphics[width=8.5cm]{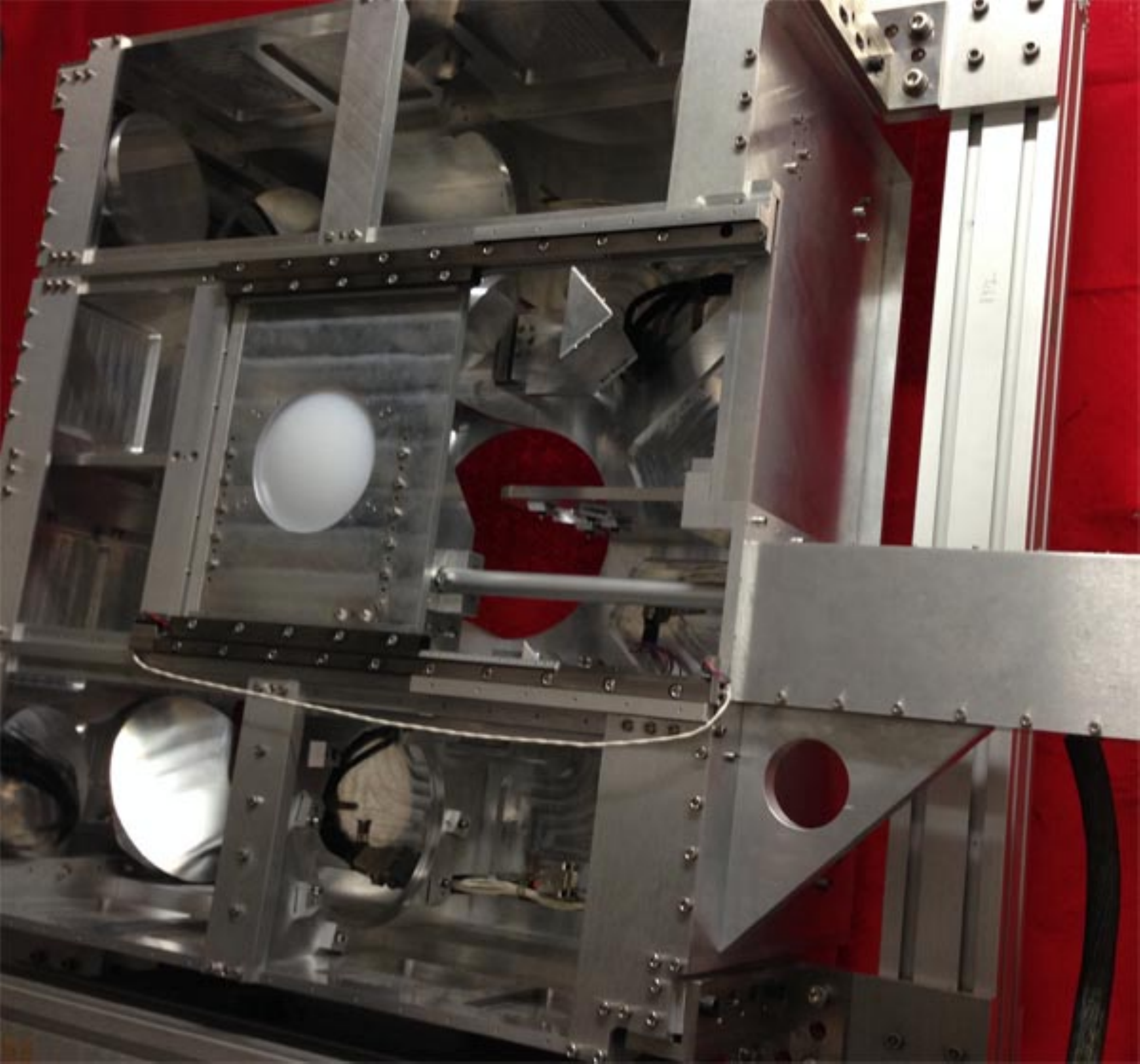}
\caption{{\bf Top:} 3D drawing of the structure supporting the 4 polarizers of the instrument (WG $_{in}$, WG  $_{out}$, WG $_{bs,L}$, WG $_{bs,R}$). {\bf Bottom:} the frame integrated with all the optical components. The volume of the main body is 70$\times$70$\times$33 cm$^3$. The protruding beam supports the linear motor that moves the folding-flat / lens assembly to exclude or include the MPI in the optical path of the OLIMPO instrument.
}\label{frame}
\end{figure}

\begin{figure}[ht]
\centering
\includegraphics[width=8.5cm]{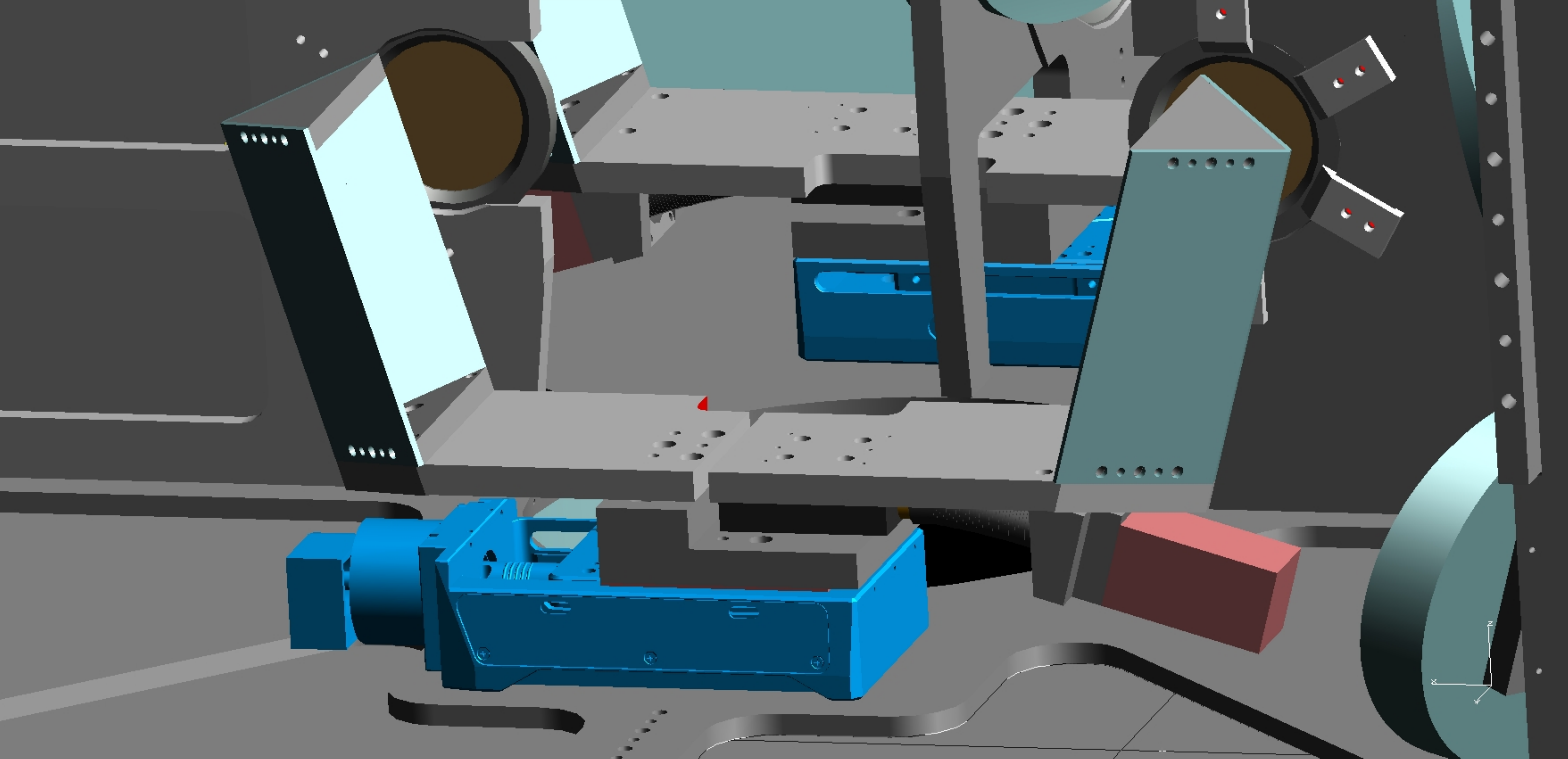}
\caption{3D drawing of the delay lines assembly, showing the two roof mirrors (one for the left interferometer and one for the right interferometer), the linear stage PLS85 that control the introduced delay (this is inserted between the delay line assembly and the optical bench), and the tuning stage APT38 (located below the right roof mirror). }
\label{delay}
\end{figure}

\begin{figure}[ht]
\centering
\includegraphics[width=3.3cm, angle=90]{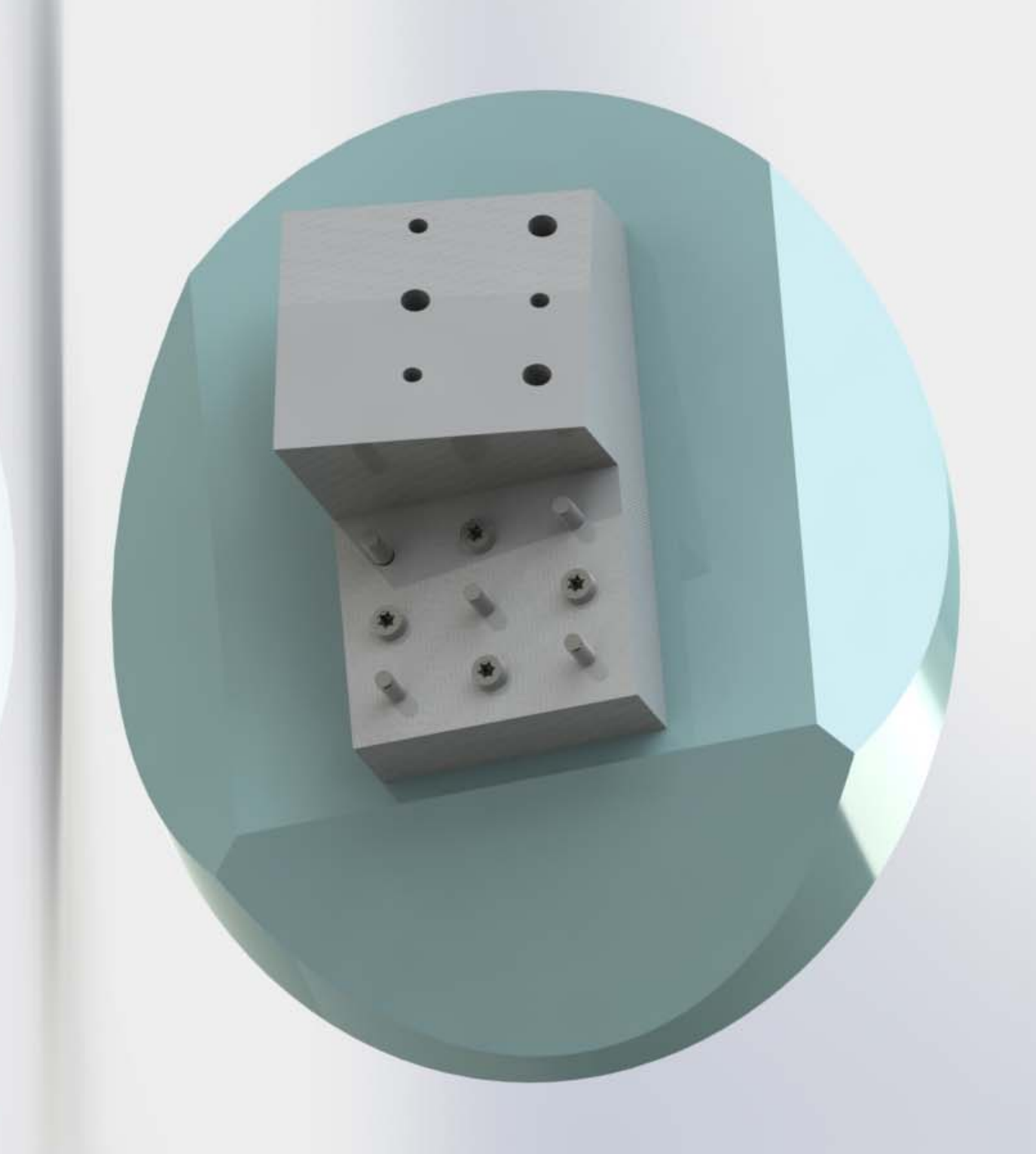}
\includegraphics[width=4.65cm]{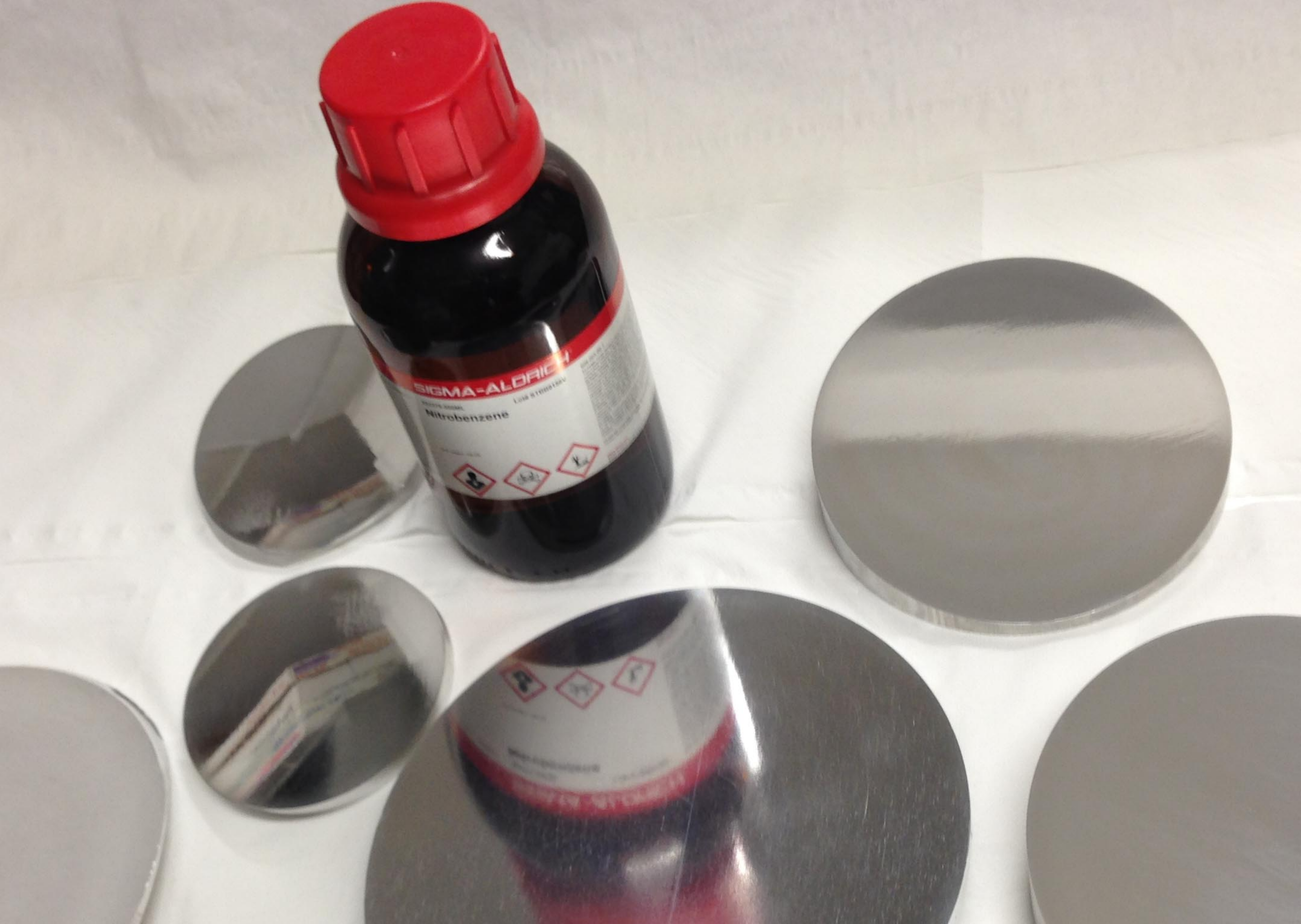}
\caption{{\bf Left:} 3D drawing of the rear side of the collimators, showing the flat interface to the optical bench. {\bf Right:} picture of the polished front surfaces. }
\label{mirror}
\end{figure}

\subsection{Mechanical structure}

The main problem in the mechanical design of a balloon-borne instrument is to create a light-weight structure stiff enough to maintain the alignment of the instrument despite the orientation with respect to the gravity vector and the considerable temperature change ($\sim$ +30C on the ground, $\sim$ -40C at float). Our constraint was to keep the weight of the spectrometer below 3\% of the total weight of the payload (i.e., 70 kg). Due to the throughput of the OLIMPO instrument, the necessary volume is large (about 150 liters). The frame supporting all the optical elements can be up to 0.7 m in size in two dimensions and needs to be squeezed to 0.3 m in the third dimension to fit the available room between the rear side of the primary mirror and the window of the detectors cryostat. We used rectified Al alloy (Mic6) flanges to build the frame. In this way, the frame and mirrors share the same thermal expansion coefficients (0.0234mm/m/C). Even if this is about a factor 2 larger than that for a steel structure, it matches the thermal expansion of the telescope, whose structure and mirrors are made of Al alloys as well. All flanges are mounted together using M5 screws and pins. Careful finite elements analysis optimization results in a relatively light-weight frame (60 kg).

We exploit the symmetry properties of our design by mounting the four beamsplitters (WG $_{in}$, WG  $_{out}$, WG $_{bs,L}$, WG $_{bs,R}$) on two crossed flanges, and all the mirror support plates near the corners of the frame. The bottom flange (10 mm thick, with large sections reduced to 3 mm)  is an optical bench on which the collimator supports and the delay lines are mounted on a side; this acts as the mechanical interface to the telescope frame on the other side. The top beams only contribute to the general stiffness of the frame and support light-weight closure sheets. The frame includes rails and linear bearings to move the folding mirror in and out of the OLIMPO optical path. All the frame parts were machined to H7 tolerance. The resulting frame is shown in Fig. \ref{frame}. The maximum deflection of the loaded optical bench under the effect of gravity (any direction) is below 0.01mm over 70 cm. To avoid internal reflections and to reduce straylight, all the internal walls of the DFTS are covered with ECCOSORB$^{\textregistered}$. Internal shields and the input hood additionally reduce the straylight level.

\subsection{Delay lines and their fine tuning}\label{delaypar}

The delay lines are the main active components in the FTS. The delay introduced must be controlled with high precision (see section \ref{prob}) and must not introduce other biases in the measurement. In our optical configuration both arms of each interferometer have moving roof mirrors. One arm is shortened while the other is elongated by the same amount. In this way, we double the optical path difference and the spectral resolution while keeping the instrument size constant. The motion of the two linear stages of the same interferometer has to be synchronized so that exactly opposite delays are introduced. We used two linear motion stages model PLS-85 from MICOS, adapted for vacuum and low-temperature operation. Each linear stage moves an assembly of two roof mirrors, one for each interferometer. The relative position of the two mirrors moved by the same stage can be tuned by means of an additional linear stage (APT38 from MICOS), which allows for fine adjustment of the distance between the two mirrors. In this way, we can finely match the optical path differences introduced by the two interferometers. The resulting moving assemblies are shown in Fig. \ref{delay}. With this arrangement, the delays in the two interferometers can be equalized to better than 0.5 $\mu m$, as checked by maximizing the zero optical path difference peak in wide-band interferograms.

\subsection{Collimator and roof mirrors}

All mirrors were machined from Al alloys (AA2024 for the flat mirrors; AA6082T6 for the curved mirrors) using CAM milling, with 0.1 mm machining steps.  These materials were selected to optimize the quality of polishing and the thermal stability. The mirror surface was obtained from a solid block including the bottom flat interface to the optical bench, as visible in Fig. \ref{mirror}. In ideal conditions, the flat surface matches the optical bench, with pins aligning the mirror in the nominal position. If this positioning would need to be modified for improved alignement, this is possible by removing the pins and using push screws.
We hand polished the surface with a cloth using polishing pastes with diamond dust grains progressively smaller (15 $\mu m$,  9$\mu m$, 3$\mu m$, and 1$\mu m$). After each polishing step the mirror was ultrasonically cleaned in ethanol. The final polishing was made using a small drill rotating a cloth dipped in 1$\mu m$ diamond dust (see Fig. \ref{mirror}). The deviation from the ideal mirror profile, as measured using a Mitutoyo long-stroke linear gage touch probe on a micrometric linear stage, is lower than 3 $\mu m$ rms for all mirrors.

\subsection{Polarizers}

\begin{figure}[!h]
  \centering
  \includegraphics[width=9cm]{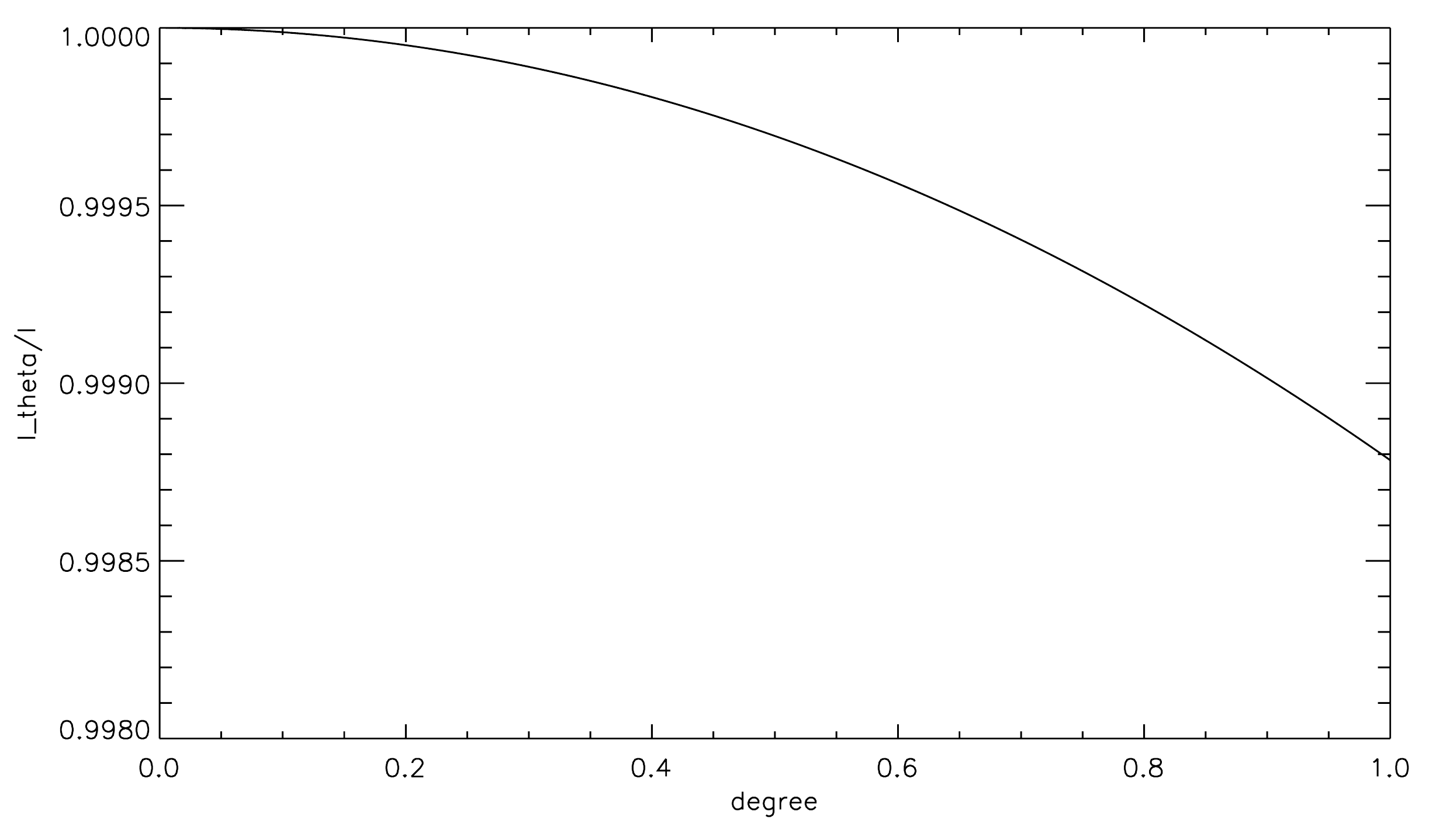}\\
  \caption{ Estimated efficiency reduction versus misalignment of the wire grid wires with respect to the optimal orientation.} \label{incorr}
\end{figure}

We used high-efficiency wire-grid polarizers, made of parallel tungsten wires, 10$\mu$m in diameter, and spaced by 25 $\mu$m for the WG$_{BS}$ and 20  $\mu$m for the WG $_{in}$ and WG  $_{out}$ , stretched across a steel support ring. This combination is suitable for large temperature excursions (is used even in cryogenic systems). The  polarizers where provided by QMC instruments and feature an efficiency very close to 100\% for both transmission and reflection, and a cross-polarization better than -40 dB.  The correct positioning of the wire grids is achieved by means of support rings with mating pins, so that the orientation of the wires is precise to better than 0.1$^o$.  The effect of an incorrect orientation of the beam-splitter wires is a modest reduction of the efficiency, as shown in Fig. \ref{incorr}. The reduction is totally negligible for a 0.1$^o$ misalignment . 

\section{Random errors and systematic effects} \label{prob}

An important problem in FTS interferometry is the accuracy and the precision of the delay line motorized stages. For a double MPI this problem can be critical, because the positioning and motion of the four roof mirrors has to be perfectly synchronous to avoid loss of signal and distortions of the interferogram.
For this system we can imagine two types of error: the first is due to random positioning errors of the roof mirrors with variance $\sigma_x$; the second one is produced by a sytematic mismatch of the delays of the two interferometers. Very precise, tunable and reproducible mechanical actuators are needed for the delay lines. The PLS-85 motorized linear stage provides 50mm of mechanical travel and a nominal positioning precision of $< 1 \mu$m rms. The APT38 equalization stages provides a resolution better than 2$\mu$m. In the following we describe the results of simulations that computed the errors introduced in the spectra by nonidealities in the delay lines.

\begin{figure}[!h]
  \centering
  \includegraphics[width=9cm]{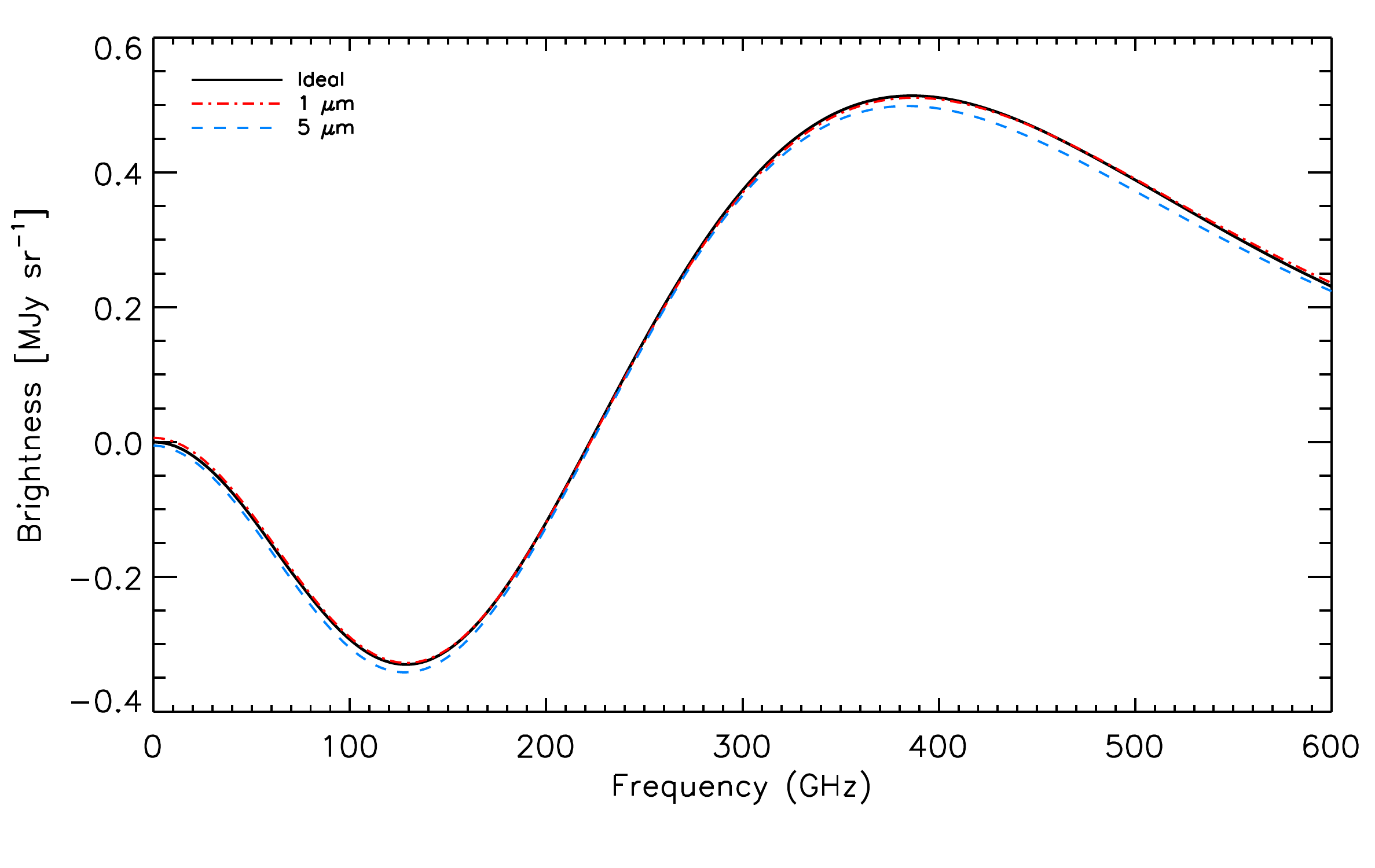}\\
  \caption{Simulated spectra of the SZE estimated from interferograms with random positioning errors
$\sigma_x$ = (0, 1, 5) $\mu$m. Even for $\sigma_x = 5 \mu m$ the systematic effect is much smaller than the random measurement erros (cfr. Fig. \ref{S+P}). } \label{random}
\end{figure}

\begin{figure}[!h]
  \centering
   \includegraphics[width=9cm]{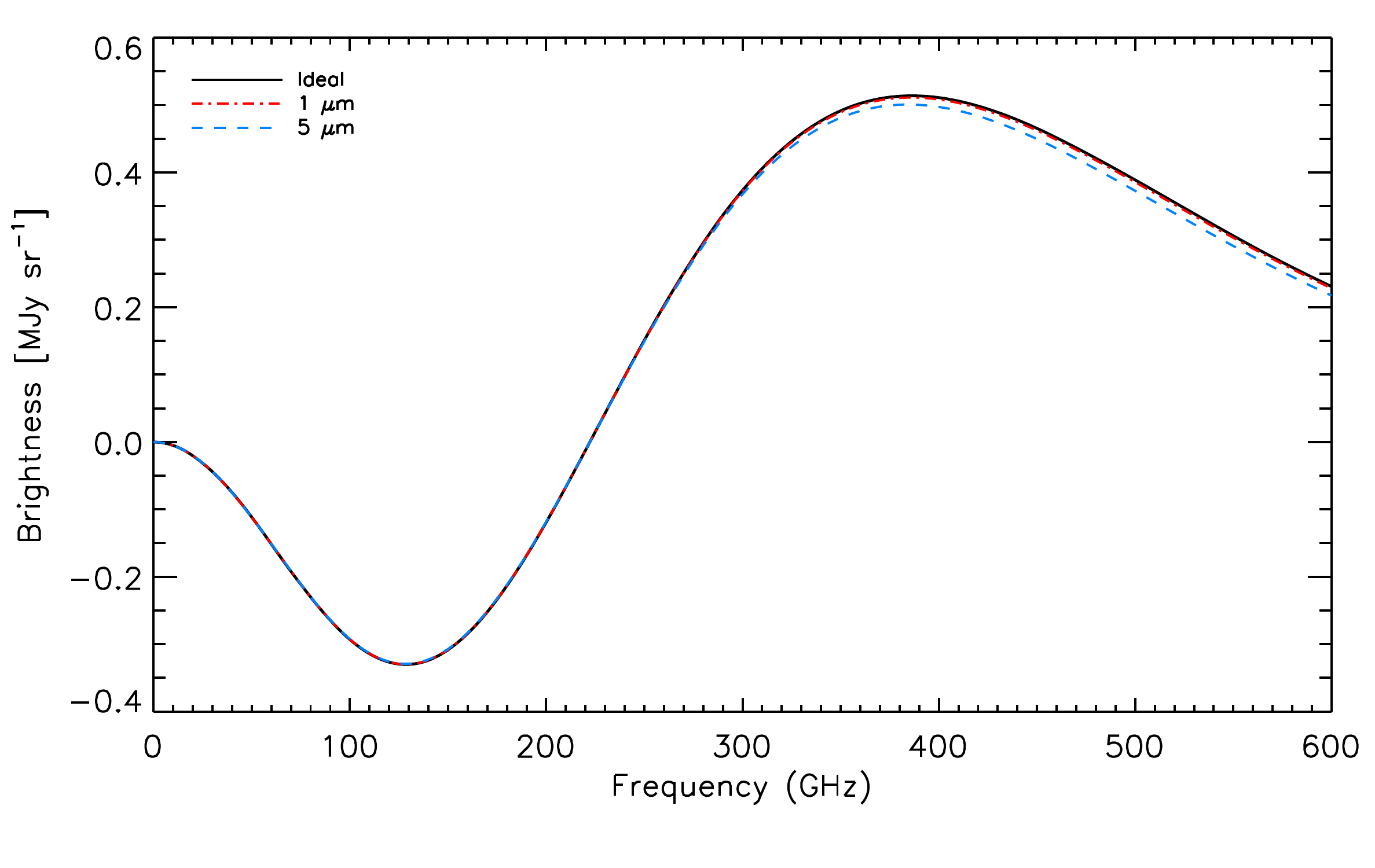}\\
 \caption{Simulated spectra of the SZE estimated from interferograms with an error in the delay line equalization of  $(0, 1, 5) \mu m$.  Again, even for $\sigma_x = 5 \mu m$ the systematic effect is much smaller than the random measurement errors (cfr. Fig. \ref{S+P}). }\label{thermal}
\end{figure}

\subsection{Positioning errors}\label{randomerrors}

We simulated interferograms measured during observations of an SZ spectrum, with random deviations from the nominal positions $x_i$. The deviations are Gaussian distributed, with a standard deviation $\sigma_x$ ranging from 1 to 5 $\mu$m.  The simulated spectra, assuming a negligible contribution from the fluctuations of the common mode, are shown in Fig. \ref{random}. The PLS-85 motorized stages used in our interferometer have a positioning accuracy $\sigma_x < 1\mu m$ rms: from the results of the simulation it is evident that this does not degrade the quality of the measurements at all.

We have also measured the total travel resulting from the sum of a large number of steps to see whether there is a measurable effect due to a possible miscalibration of the step size. For both stages the step size is accurate to 0.01\%, so any shift of the frequency scale in the measured spectrum will be $\simlt 0.01 \%$. Given the medium resolution of this instrument, this is negligible.

\subsection{Tuning of the delay line lenghts}

The two interferometers should process radiation by introducing exactly the same delay. For this reason, as explained in Sect. \ref{delaypar}, we added a tuning stage APT38 to match the delays in the two interferometers. A balloon-borne instrument changes its temperature by a significant amount, from about +30C before launch and in the laboratory to $\sim$ -40C at float. The resulting change in all dimensions of the FTS could produce a shift of the delay lines positions with respect to the laboratory ones. For this reason, the APT38 can be moved remotely, during the flight, by means of telemetry commands. We performed a simulation to estimate the effect of a mismatch of the two delay lines ranging from 1 to 5 $\mu m$ for SZ observations, as in Sect. \ref{randomerrors}. The results are shown in Fig. \ref{thermal}. With 1-2 $\mu$m of mismatch, achievable with the APT38, the effect is a very slight deformation at the high-frequency end of the measured spectrum,  much smaller than the expected random errors due to photon and detector noise (cfr. Fig. \ref{S+P}). Using the same parameter extraction procedure as described in section \ref{ps}, we verified that the shift in the best-fit values of the parameters induced by this effect is smaller than their 1$\sigma$ error. 

The tuning lines can be operated by telecommand during the flight of the instrument. For an accurate equalization of the two FTSs, several interferograms with different equalization delays will be taken while observing a planet, for example, Mars. The signal-to-noise ratio in this case is very high (thousands of times higher than for SZ signals, with the OLIMPO telescope) so that in less than one hour of measurements the equalization settings that maximize the amplitude of the interferograms can be found.

\section{Tests and commissioning} \label{test}

\subsection{Alignment }

With all mirrors in their nominal positions, the wire grids were replaced by semi-reflective optical glasses (3 mm thick), which provide beam-splitting of visible light from an expanded laser placed at the input port of the spectrometer. A target cross-wire is present in the optical glasses, so that collimators for each beamsplitter can be adjusted to center the laser beam. We are able to trace the whole optical path of the MPI checking the output positions of all the fields. The required mechanichal adjustments were minimal, confirming the machining accuracy.
After optical alignment, we started to measure interferograms of mm/submm sources.

\subsection{Monochromatic measurements using a Gunn diode}

\begin{figure}[ht]
\centering
\includegraphics[width=5.0cm, angle=90]{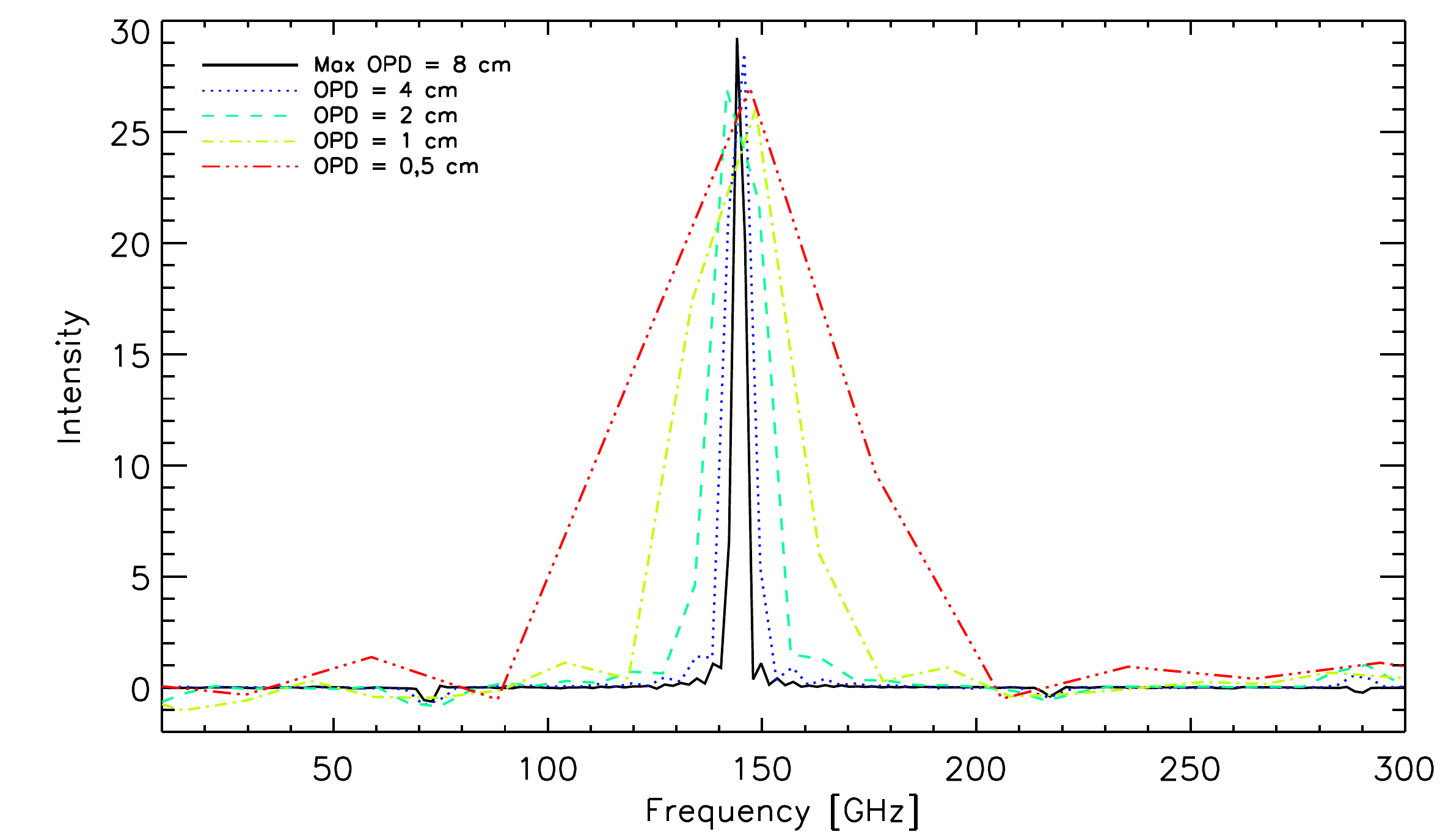}
\caption{
Spectra of Gunn diode emission obtained for different maximum optical path differences. Since the Gunn diode is effectively monochromatic, the measured widths represent direct measurements of the spectral resolution of the DFTS.} \label{gunn}
\end{figure}

A 140 GHz Gunn diode with a power of a 20-30 mW was used with a Golay cell room-temperature detector to measure the commissioning interferograms. The diode output is basically monochromatic (the first harmonic is down by a factor $\sim$ 20 and we used a low-pass filter with cut-off at 220 GHz to further reject the harmonics) and was coupled to the interferometer using a single-mode feedhorn and two high-density poliethilene lenses, to match the input $f/\#$ of the DFTS. In Fig. \ref{gunn} we present the interferogram obtained moving one roof mirror up to the maximum stroke of +2cm from the zero path difference (ZPD), and the other symmetrically to a maximum offset -2cm from the ZPD, producing a maximum optical path difference (OPD) of 8 cm,  and spectra obtained for different maximum optical path differences. For the largest possible optical path difference introduced by our DFTS (8 cm) the width of the 140 GHz line is 1.87 GHz, consistent with the expectation $\Delta \sigma \sim 1/(2 \ OPD_{max})$ (the interferogram was apodized with a triangular window). Decreasing the OPD, the spectral resolution decreases as expected, while the signal-to-noise ratio for the measurement of continuum sources improves as $1/\Delta \sigma$.

\subsection{Continuum measurements}

To test the performance at higher frequency, we coupled the DFTS to a thermal source, an Hg lamp (Philips 125W) producing a $\sim$ 5000K gray body, and filtered the emission using different mesh filters. The source was coupled to the interferometer by means of a two-lenses system, to reproduce the same f/\# and beam as the OLIMPO telescope.

\begin{figure}[ht]
\centering
\includegraphics[width=5.0cm, angle=90]{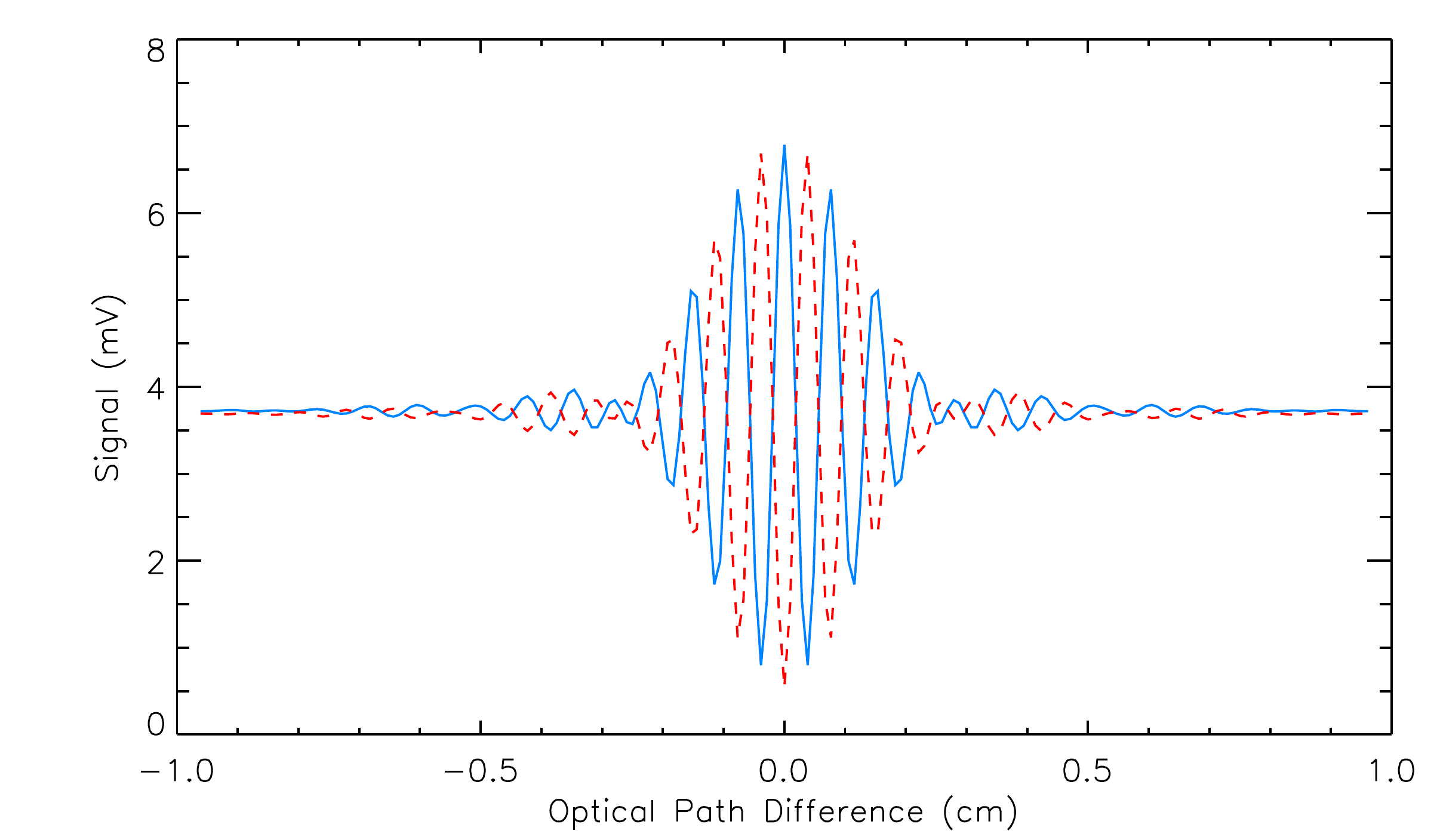}
\includegraphics[width=8.5cm]{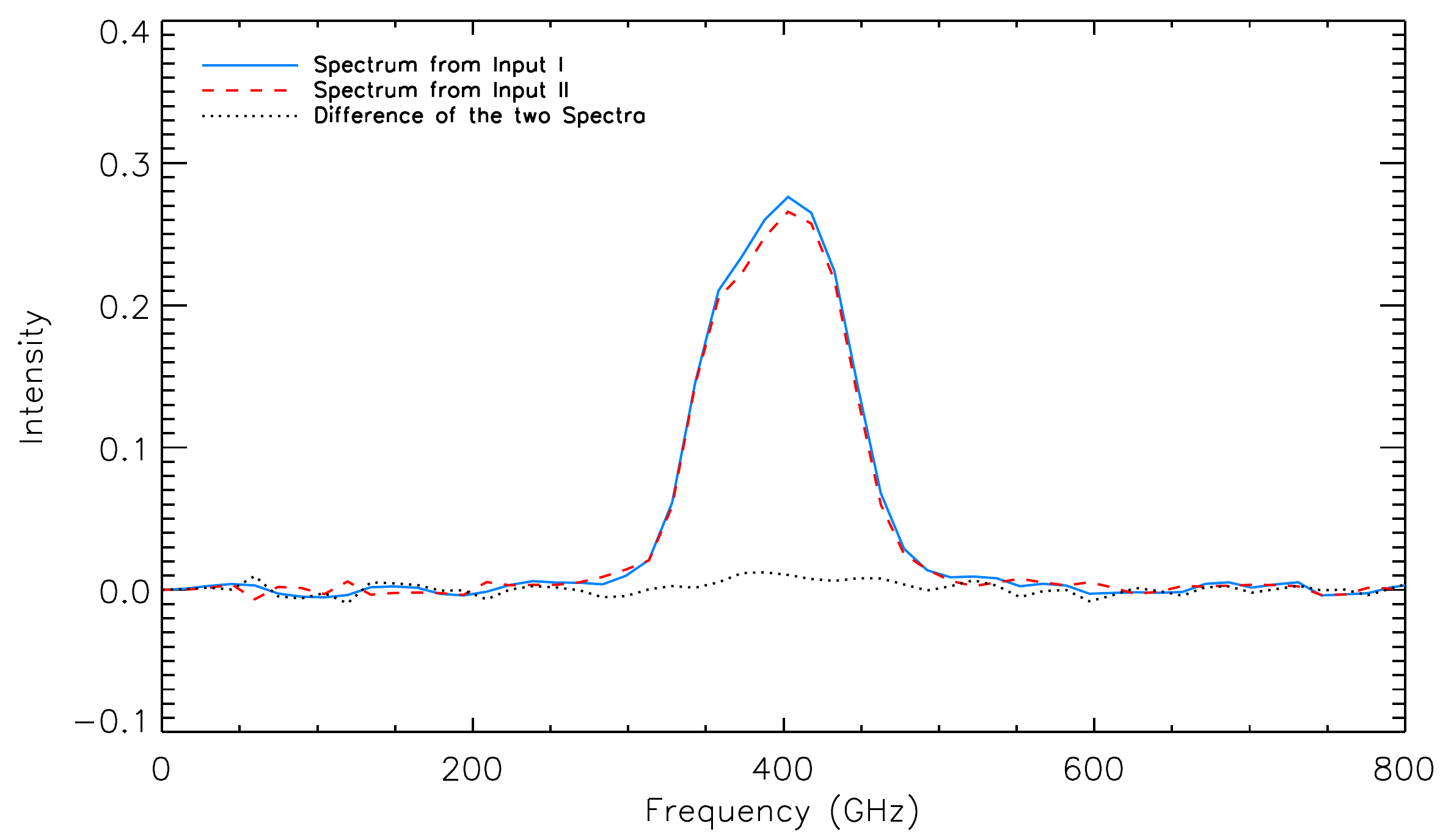}
\caption{{\bf Top:} Interferograms measured with an Hg discharge lamp and a room-temperature blackbody, band-pass filtered at 375 GHz. In the continuous interferogram, the Hg lamp is placed in the center of port A, while the center of port B is filled with the room-temperature blackbody. In the dashed interferogram the positions of the two sources are switched. {\bf Bottom:} spectra obtained from the two interferograms above; the second spectrum has been multiplied by -1 to facilitate the comparison with the first one. }\label{350GHz}
\end{figure}

In Fig. \ref{350GHz} we plot the interferograms obtained through a 375 GHz bandpass filter for two symmetric positions of the source, focusing radiation from the Hg lamp on the centers of the two input ports of the DFTS (which are the centers of the two surfaces of the input wedge mirror). In the first interferogram ($I_1$), the center position of port A is filled by the Hg lamp, while the center position of port B is filled with a room-temperature blackbody. In the second one ($I_2$), the two sources are switched, so that the Hg lamp is located in the center position of port B. The two measured interferograms feature a good symmetry, as expected for a truly differential instrument. We derive $I_2 \simeq -1.03 I_1$. The $\sim 3\%$ asymmetry mainly arises because the Hg lamp does not fill the beam, and our repositioning accuracy is about 0.2 mm with a distance between the two symmetrical positions of 26 mm and a size of the lamp aperture of 8 mm; moreover, the gain stability of the Golay cell and of the Hg lamp emission are at the \% level, on the timescales involved in this measurement, as checked by repeating this procedure several times. Therefore, while confirming the differential nature of the instrument, this result cannot be used to characterize the common-mode rejection ratio $CMRR$ of our DFTS. This was measured by filling the two ports with two room-temperature blackbodies with different temperature, measuring the resulting interferogram $I_1$, then switching the two blackbodies and measuring the resulting interferogram $I_2$ . The simplest model for an asymmetry in the interferometer is described by a modified equation \ref{5}:
\begin{equation}\label{5m}
I_L = \frac{1}{2}(I_a+I_b)+\frac{1}{2}\left[I_a - (1-CMRR) I_b\right]\cos(\delta) \ .
\end{equation}
From our measurements of $I_1$ and $I_2$ we find $CMRR =(8 \pm 6)\times 10^{-4}$, that is, we can set an upper limit $CMRR < 2 \times 10^{-3}$ at 95\% C.L. .

\begin{figure}[ht]
\centering
\includegraphics[width=5.2cm, angle=90]{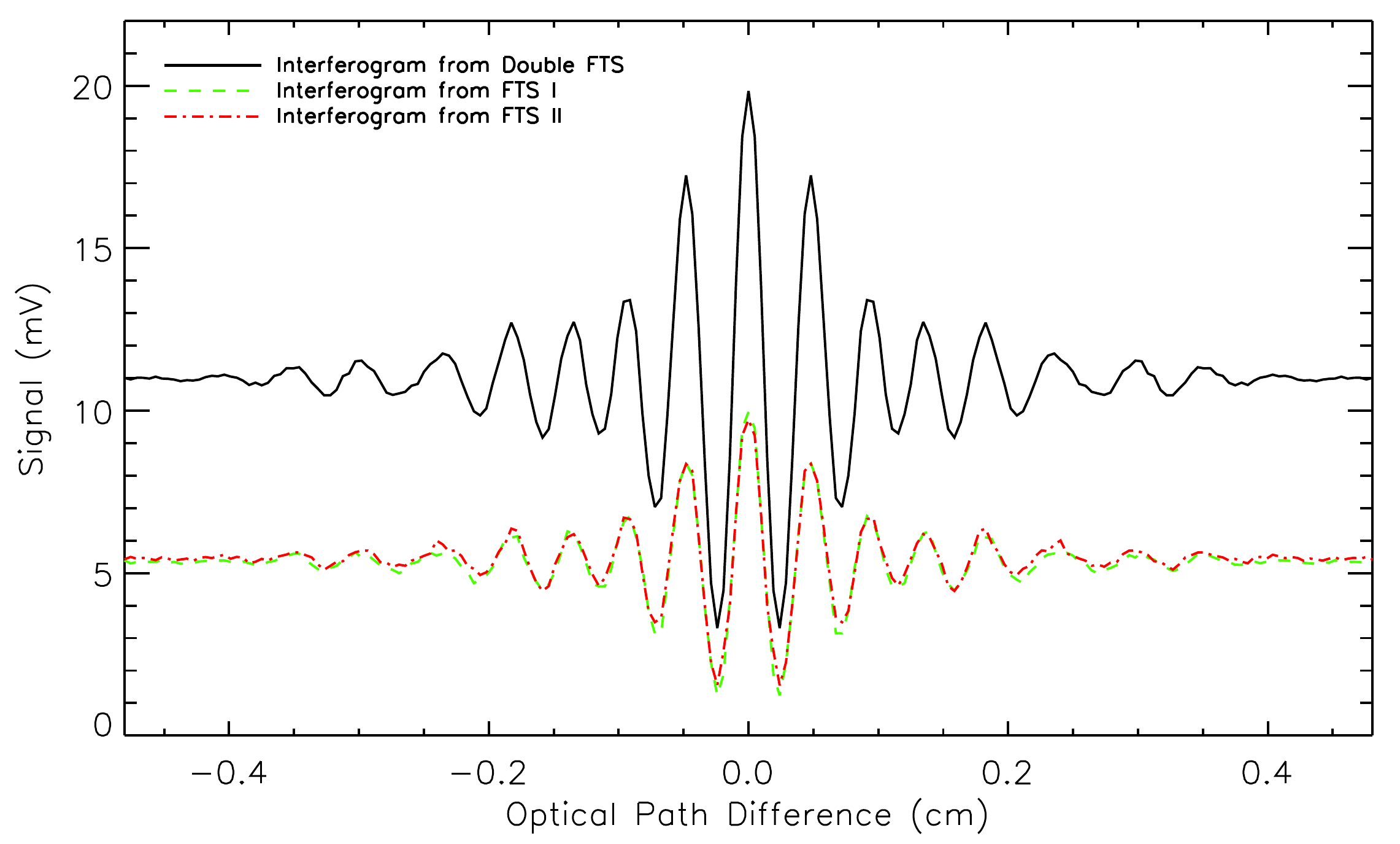}
\caption{Interferograms measured with an Hg discharge lamp, band-pass filtered at 600 GHz, for operation of the left FTS alone, of the right FTS alone, and of the complete DFTS. The two FTSs contribute almost equally to the total signal, and the DFTS is still modulating efficiently at this high frequency. }\label{2fts}
\end{figure}

In Fig. \ref{2fts} we compare the interferograms produced by each of the two FTSs present in the DFTS (each obtained by blanking the WG$_{BS}$ of the other FTS) with the interferogram obtained from the simultaneous operation of the two FTSs. In this case, a 600 GHz band-pass filter was used to reduce the spectral width of the source. Even at this high frequency (basically the highest operative frequency of the instrument), the DFTS efficiently modulates the radiation under analysis (see section \ref{effi}). We found that the quality of the equalization of the two interferometers is totally dominated by the aligment of the mirrors and wire grids in the DFTS. In the case shown in Fig. \ref{2fts}, the difference between the two interferograms is mainly in the baseline level (again due to the instability of the Golay cell and of the Hg lamp); the rest of the difference (about 0.7\%) is very well described by an overall constant efficiency factor. In addition to this effect, which is corrected for by the overall in-flight calibration procedure (see next section for a discussion), the two FTSs contribute almost equally to the total signal.

\subsection{Efficiency of the instrument} \label{effi}

Our instrument can be bypassed to perform photometric measurements, as explained in section \ref{plug-in}. When inserted, it reproduces at the output the same beamwidth and size as the input beam.  Comparing the ZPD signal transmitted by our instrument to the ZPD signal measured with the same detector when the spectrometer is bypassed, we can assess the optical efficiency of the instrument. We find that this is dominated by the alignment of the mirrors and wiregrids in the interferometers. After careful alignement, we measured that  for unpolarized sources the ratio between the two signals is $\simgt 90\%$ at 150 GHz, $\simgt 95\%$ at 220 GHz and 375 GHz, and $\simgt 80\%$ at 600 GHz. Given the sensitivity of these numbers to alignement of the mirrors, and the fact that when the alignment is not perfect, the efficiencies above are all reduced by the same factor, we conclude that the slight inefficiency is mainly due to edge rays that are lost in case of inaccurate alignment. 

In the spectroscopic configuration we compared the height of the ZPD peak (above the baseline) with the baseline amplitude to assess the interferometric modulation efficiency of the instrument. This is $\simgt 98\%$ at 145 GHz and $\simgt 80\%$ at 600 GHz, demonstrating that the instrument features a good efficiency in the whole frequency range useful for SZ measurements. The reduction of the modulation efficiency at high frequency is probably due to same small misalignments, mispositioning, and machining imperfections of the mirrors that cause the efficiency reduction described above. In this case a small fraction of the beam undergoes a slightly different geometrical path that the bulk of the rays, which generates randomly distributed phase shifts. Constructive interference will be affected more at short wavelengths than at long wavelengths. These effects are factorized with the spectral response of the sub-band filters and of the detectors, and must all be calibrated during the flight by observing a reference Rayleigh-Jeans source, measuring the relative spectral response of the whole instrument.

\section{Discussion and conclusion} \label{disc}

We have developed a differential Fourier-transform spectrometer optimized for low-medium-resolution spectroscopy of faint brightness gradients. The instrument is efficient, features large throughput ($\sim  2.3 \ cm^2 sr$) and medium resolution ($\sim 1.9$ GHz FWHM, independent of frequency), in a relatively compact setup. The instrument is suitable for suborbital operation and will be tested in the forthcoming flight of the OLIMPO stratospheric balloon. 

We tested several possible systematic effects. The only one resulting in a potentially non-negligible systematics is the $CMMR$, for which we set an upper limit $CMMR < 0.2\%$. For measurements of the SZ effect with OLIMPO, where the DFTS works at room temperature, the common-mode background on the detectors is due to the instrument itself (mainly the room-temperature mirrors and the WGs), the residual atmosphere, the CMB, and interstellar dust. The resulting common-mode brightness, multiplied by our CMRR upper limit, is about (6, 20, 60, 160) MJy/sr in our four bands, to be compared to SZ signals about (-0.2, 0.2, 0.7, 0.8) MJy/sr. An offset removal observation strategy is evidently in order. The SZ measurement will be performed by subtracting from a first observation of the brightness difference between the cluster and the reference direction, a second observation of the brightness difference between two other nearby blank-sky directions. With a stability of the common-mode emissions better than 0.05\%, this strategy allows good measurements of the SZ. We stress that what we just described is the worst case allowed by our upper limit on the CMRR. Our instrument model suggests a much better situation. The confirmation will come from in-flight measurements.

A second implementation of the DFTS was designed for operation in the W-band at the Sardinia radio telescope. 

With a different mechanism for the mirror motion (see e.g. Schillaci et al. \cite{Schi11} and references therein), the system can be implemented for operation in a cryogenic environment, suitable for space-based operation. This is the case of the SAGACE (de Bernardis et al. \cite{debe10}) and millimetron (Smirnov et al. \cite{Smir12}) missions. Here the CMRR offset due to instrument emission will be reduced by at least a factor 100, resulting in very competitive spectrometers for wide-band continuum measurements.

\acknowledgements This activity has been supported by MIUR PRIN 2009 {\sl "Mm and sub-mm spectroscopy for high resolution studies of primeval galaxies and clusters of galaxies"} and by the contract {\sl OLIMPO} of the Italian Space Agency. We thank Giorgio Amico and Stefano Banfi for carefully machining several parts of the instrument. We thank Sebastiano Spinelli for implementing the procedure and test set-up for surface roughness and profile measurements on mirrors. We thank Marco De Petris for help in interfacing the DFTS to laboratory sources. C. P. Novaes acknowledges the CNPq [237059/2012-6] fellowship.

\end{document}